\documentclass[%
 reprint,
superscriptaddress,
 amsmath,amssymb,
 aps,
]{revtex4-2}

\usepackage{graphicx}% Include figure files
\usepackage{dcolumn}% Align table columns on decimal point
\usepackage{bm}% bold math
\usepackage{xcolor}
\usepackage{braket}
\usepackage{multirow}
\usepackage{array}
\usepackage{diagbox}
\usepackage[hidelinks]{hyperref} 
\newtheorem{theorem}{Theorem}

\newtheorem{corollary}{Corollary}
\usepackage{amsmath}
\usepackage{xr-hyper}
\usepackage{soul}
\usepackage{titletoc}

\newcommand{\QuICS}{Joint Center for Quantum Information and Computer Science, National Institute of Standards and Technology and
 University of Maryland, College Park, Maryland 20742, USA}
\newcommand{\JQI}{Joint Quantum Institute, National Institute of Standards and Technology and University of Maryland, College Park, Maryland 20742, USA}
 \newcommand{\QINC}{A*STAR Quantum Innovation Centre(Q.InC), Agency for Science,
Technology and Research(A*STAR), 2 Fusionopolis Way, 08-03 Innovis 138634, Singapore}
\newcommand{\UTS}{%
Centre for Quantum Software and Information, University of Technology Sydney,\\
Sydney, New South Wales 2007, Australia%
}
\newcommand{\ANU}{Department of Quantum Science and Technology, Research School of Physics,  The Australian National University, Canberra, ACT 2601, Australia.}
\newcommand{\CQT}{Centre for Quantum Technologies, National University of Singapore,
117543, Singapore}
 
\begin{document}
\setcounter{tocdepth}{-1}
\preprint{APS/123-QED}

\title{Quantum advantage in learning single mode bosonic channels}
%\title{Exponential Learning advantage in single bosonic mode}
% Force line breaks with \\

\author{Angela Anna Baiju}
\email{angelbaiju08@gmail.com}
\affiliation{\QINC}

%\affiliation{%

%\affiliation{
 % with \\
%}%
%\affiliation{
 
%}%

%\affiliation{%
%}%


\author{Aritra Das}
\affiliation{\UTS}
\author{\"{O}zlem Erk{\i}l{\i}\c{c}}
\affiliation{\UTS}
\author{Jiayi Qin}
\affiliation{\ANU}
\author{Li Gong}
\affiliation{\QINC}
\author{Syed M Assad}
\affiliation{\QINC}
\author{Ping Koy Lam}
\affiliation{\QINC}
\affiliation{\CQT}
\author{Biveen Shajilal}
\email{Current affiliation: Q-CTRL, Sydney, Australia.}
\affiliation{\QINC}
\author{Lorc{\'a}n O Conlon}
\email{lorcanconlon@gmail.com}
\affiliation{\JQI}
\affiliation{\QuICS}
\author{Jie Zhao}
\email{jie.zhao@anu.edu.au}
\affiliation{\ANU}

\newcommand{\ZJ}[1]{{\color{black}#1}}


\begin{abstract}
%Quantum resources can substantially reduce the amount of data required to learn physical systems, with exponential improvements in sample complexity emerging in several quantum learning tasks. Yet these advantages have typically been demonstrated in settings where the required quantum resource grows with the complexity of the problem, most notably through entanglement or increasing system dimension. This raises a fundamental question: can an exponential learning advantage instead be obtained within a fixed, unentangled quantum system? We \ZJ{address} this question \ZJ{where we consider learning} an unknown random-displacement distribution whose complexity is determined not by the dimensionality of the physical system, but by the Fourier resolution of its features. \ZJ{We show that quantum limited noise associated with vacuum probes} progressively obscures high-frequency features, imposing an exponential sample complexity cost. We establish an information-theoretic lower bound for arbitrary classical-state probes and show that squeezing circumvents \ZJ{the classical limit} by extending the accessible Fourier bandwidth. Experimentally, we \ZJ{demonstrate} exponential reduction in sample complexity \ZJ{using squeezed vacuum probes} in both binary hypothesis testing and characteristic-function reconstruction. Our results show that \ZJ{a single bosonic mode} can support exponential quantum learning advantages without entanglement or an increase in system size, identifying accessible Fourier bandwidth as a distinct resource for quantum learning.
Quantum resources can dramatically reduce the data required to learn physical systems, with exponential improvements in sample complexity demonstrated in several quantum learning tasks. However, these advantages have typically relied on quantum resources that scale with problem complexity, most notably increasing system dimension or entanglement. This raises a fundamental question: can exponential quantum learning advantages arise within a fixed, unentangled quantum system? In this paper, we address this question by considering the learning of an unknown random-displacement distribution whose complexity is determined not by the dimensionality of the physical system, but by the Fourier resolution of its features. We show that the quantum-limited noise of vacuum probes progressively obscures high-frequency features, leading to an exponential growth in sample complexity. We establish an information-theoretic lower bound for arbitrary classical-state probes and show that squeezing overcomes this classical limit by extending the accessible Fourier bandwidth. Experimentally, we demonstrate an exponential reduction in sample complexity using squeezed vacuum probes for both binary hypothesis testing and characteristic-function reconstruction. Our results show that a single bosonic mode can exhibit exponential quantum learning advantages without entanglement or an increase in system size, identifying accessible Fourier bandwidth as a fundamentally distinct resource for quantum-enhanced learning.
\end{abstract}

\maketitle

%\tableofcontents
\section{\label{sec:level1}INTRODUCTION}

Inferring the properties of an unknown quantum system from measurement data lies at the heart of quantum information processing. Whether the goal is to characterize an unknown quantum process, infer hidden parameters or reconstruct an underlying probability distribution, the performance of an inference protocol can be characterized by the number of observations required to achieve a prescribed accuracy. The fundamental sample complexity limits for such inference tasks from a probability distribution are well established ~\cite{vapnik1998statistical}. The challenge of learning from quantum systems is further sharpened by fundamental uncertainty principles, which state that quantum measurements inevitably and irreversibly disturb the measured states \cite{conlon2026100}. \ZJ{Interestingly}, recent works have shown that \ZJ{such limits can be circumvented by} utilising quantum resources at the probe preparation, processing, and measurement stages, \ZJ{yielding substantial complexity} reduction \cite{kim2026fundamental, huang2021information,lee2023evaluating}. 
\begin{figure*}[t]
    \centering
    \includegraphics[width=\textwidth]{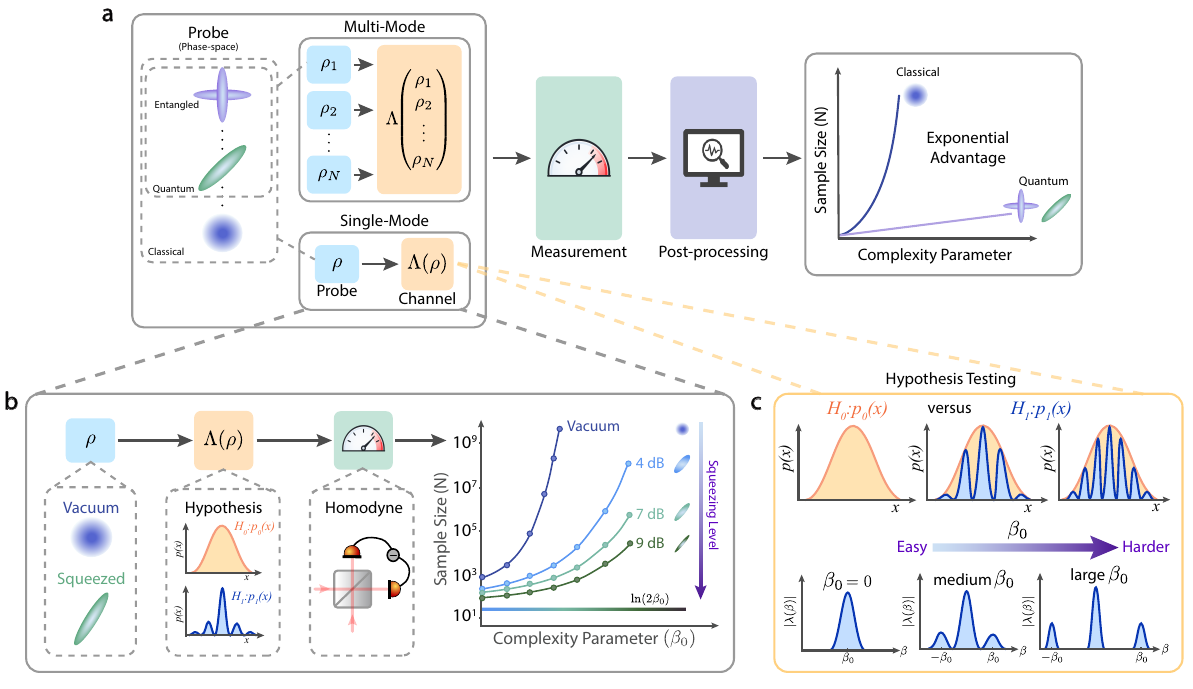}
    \caption{\textbf{Exponential quantum learning advantage 
    from a single squeezed mode.}
    \textbf{(a)}~General framework for quantum channel 
    learning. A probe state is sent through an unknown 
    channel $\Lambda$, measured, and the outcomes are 
    post-processed to infer the channel's properties. The 
    choice of probe states\ZJ{—}classical, squeezed, or 
    entangled \ZJ{states—}and the measurement strategy jointly 
    determine the sample complexity: the number of channel 
    uses required to achieve a prescribed learning accuracy. 
    Prior work has established that multi-mode entangled 
    probes can reduce sample complexity exponentially 
    compared to any classical strategy (right panel), but 
    has left open whether such an advantage \ZJ{necessitates} 
    entanglement or many modes.
    \textbf{(b)}~Our setting: a single bosonic mode $\rho$, \ZJ{such as}
    vacuum or squeezed probe state,  is sent through 
    a random displacement channel $\Lambda(\rho)$ and 
    detected by homodyne measurement. The learning task 
    is to distinguish between two displacement 
    distributions, a single Gaussian ($H_0$) and an oscillatory Gaussian distribution ($H_1$) whose characteristic function contains three Gaussian peaks. The right panel shows that at any fixed 
    squeezing level, a squeezed probe exponentially outperforms any 
    vacuum or coherent-state probe with homodyne measurement, with the advantage growing without
    bound as the task difficulty increases, establishing an exponential 
    quantum learning advantage from a single mode and without entanglement. Above a critical squeezing level, this reduction is sufficient for the squeezed-probe complexity to fall exponentially far below the classical lower bound.
    \ZJ{Remarkably, by} letting the squeezing grow
    with the task difficulty, $r\sim\ln2\beta_0$, \ZJ{a constant complexity $\mathcal{O}(1)$ is achievable} even as $\beta_0$ grows asymptotically large \ZJ{(flat line)}.
    \textbf{(c)}~\ZJ{Hypothesis testing}. The 
    two hypotheses differ only in the separation $\beta_0$ 
    of the side peaks in characteristic function space \ZJ{(bottom pannel)}, 
    corresponding to increasingly fine oscillatory 
    structure in the displacement distribution \ZJ{(top panel)}. As $\beta_0$ grows the distinguishing 
    features become finer and the learning task 
    exponentially harder for any fixed-noise probe. 
    Squeezing reduces the probe noise, extending the 
    \ZJ{accessible bandwidth determined by $\beta_0$,} and preserving 
    sensitivity to fine features that vacuum probing 
    cannot resolve.}
    \label{fig:fig1}
\end{figure*}

A major development in this field has been that quantum entanglement can provide an exponential reduction in sample complexity for certain tasks. In discrete-variable (DV) systems, quantum memory, ancillary qubits, and entangled measurements enable exponentially more efficient learning of quantum states, channels, and dynamical processes than any classical protocol\ZJ{s}~\cite{huang2022quantum, 
chen2022quantum, liu2024exponential, seif2024entanglement, kim2026fundamental}. Extending this to continuous-variable (CV) systems, Oh et al.~\cite{oh2024entanglement} proved that learning a bosonic random displacement channel requires a number of samples exponential in the number of modes $n$ without entanglement, while an entangled probe can complete the same task with sample complexity independent of \ZJ{$n$} for sufficiently large squeezing. This was recently demonstrated experimentally at scale, yielding reductions exceeding eleven orders of magnitude over coherent-state probes~\cite{liu2025quantum}. These results establish entanglement as a powerful resource for overcoming sample-complexity barriers.  At the same time, the above works also share a structural feature: the complexity of the learning task is coupled to the physical size of the system. In DV systems this coupling is natural as the Hilbert space grows with the number of qubits, and so does the complexity of the typical learning tasks \cite{kim2026fundamental, huang2021information,lee2023evaluating,huang2022quantum, 
chen2022quantum, liu2024exponential, seif2024entanglement}. Kim et al. \cite{kim2026fundamental} \ZJ{showed} that exponential learning advantages can persist with vanishingly small entanglement, provided that the quantum memory has sufficiently large dimension. This distinction, between entanglement and dimension as a resource is particularly relevant to CV systems, where even a single bosonic mode occupies an infinite-dimensional Hilbert space, offering a richer \ZJ{encoding} space than a single qubit without requiring additional modes. Recent works have established fundamental sample-complexity bounds for learning continuous-variable quantum states through quantum-state tomography, highlighting the role of system size and energy constraints in CV learning~\cite{mele2025learning,chen2026towards,mele2026advances}. This raises a fundamental question: can an exponential learning advantage exist without entanglement, or can such an advantage arise from other quantum resources when the learning complexity is decoupled from system size?

A natural precedent from CV quantum metrology suggests that an exponential learning advantage \ZJ{may arise} without entanglement \cite{kwon2019nonclassicality}. In standard parameter estimation, squeezed states concentrate quantum noise into one quadrature of the electromagnetic field, \ZJ{enabling a} polynomial enhancement in precision \ZJ{beyond the shot noise limit set by vacuum fluctuations}\cite{caves1981quantum,bradshaw2018ultimate}. \ZJ{Analogously,} in a learning problem, a well engineered probe state can extract information \ZJ{more efficiently} (\ZJ{see} Fig.~\ref{fig:fig1}a). Vacuum and coherent probes are bandwidth-limited by their quantum noise, exponentially suppressing fine-grained features of the unknown signal. Squeezed states extend this bandwidth by redistributing quantum noise, allowing the same features to be recovered with exponentially fewer measurements. Crucially, this bandwidth improvement changes the exponent in sample complexity rather than a constant prefactor, turning what is a polynomial metrological gain into an exponential learning advantage without entanglement.
Here we demonstrate this explicitly for the task of learning a random, single-mode, single-quadrature displacement channel $\Lambda$ (Fig.~\ref{fig:fig1}b) where the complexity of the task is controlled by the resolution parameter $\beta_0$, rather than by system dimension. The task is formulated as binary hypothesis testing between a single Gaussian distribution ($H_0$) and an oscillatory Gaussian distribution ($H_1$), whose characteristic function contains three Gaussian peaks. The parameter $\beta_0$ quantifies the separation between the two side peaks in Fourier space \ZJ{(Fig.~\ref{fig:fig1}c)}.
This is a special case of \ZJ{the} channel \ZJ{studied} in Ref.~\cite{oh2024entanglement}. \ZJ{The} problem is operationally natural: it models the inference of fine-grained phase-space structure from noisy \ZJ{Gaussian} measurements, directly relevant to quantum sensing, spectroscopy, and \ZJ{Gaussian} channel characterization \cite{fadel2025quantum,lvovsky2009continuous}. The two hypotheses differ only in the location of their distinguishing feature in characteristic function space, which can be tuned continuously to higher Fourier frequencies corresponding to finer spatial structure in the phase space distribution as shown in Fig.~\ref{fig:fig1}c. As this feature frequency increases, the probe state's characteristic function acts as an exponentially decaying filter, suppressing the distinguishing signal. Any probe state that extends the accessible measurement bandwidth by reducing its effective noise variance therefore yields an exponential reduction in sample complexity. A squeezed vacuum probe achieves precisely this: by redistributing quantum noise it extends the accessible bandwidth, yielding a sample complexity exponentially smaller than any vacuum or coherent state probe\ZJ{s} for both binary hypothesis testing and full reconstruction of distributions $H_0$ and $H_1$. The exponential speedup is therefore not inherently tied to entanglement, but to the ability to control estimator variance through appropriate state preparation — an insight that suggests entanglement-free quantum learning advantages in structured single-mode inference problems, realizable with current CV technology using only single mode squeezing and homodyne detection.

We  provide a rigorous information-theoretic lower bound showing that every vacuum-state protocol requires a number of samples growing exponentially with $\beta_0$, independent of the estimator, adaptive strategy, or post-processing employed. We demonstrate experimentally that squeezed-state homodyne protocols accomplish the task using exponentially fewer samples for both binary hypothesis testing and characteristic-function reconstruction. These results establish a distinct mechanism for exponential quantum 
learning: unlike previous approaches based on entanglement or quantum memory, the advantage demonstrated here arises entirely from single-mode squeezing. More broadly, our work identifies the accessible Fourier bandwidth as a fundamental resource for CV quantum learning\ZJ{—}exponential quantum 
advantages can emerge from appropriately structured inference problems \ZJ{without requiring} entanglement.

\section{\label{sec:theory}Single  Mode Quantum Learning}
This section outlines the theoretical framework for learning a single-mode random displacement channel and derive the sample-complexity scaling with the Fourier-space resolution parameter $\beta_0$.

\subsection{Channel Model and Inference Tasks}

Consider a single-mode random displacement channel along the $x$-quadrature, defined by
\begin{equation}
\Lambda(\rho) = \int dx \ p(x)\, D(x)\rho D^\dagger(x).\
\end{equation}
where $D(x) = e^{-ix\hat{p}/2}$ is the displacement operator along $\hat{x}$ such that $\hat{x}\mapsto\hat{x}+x$ and $p(x)$ is the unknown classical probability distribution of displacements. The channel can be fully characterized by its displacement characteristic function \ZJ{{\it i.e.} the Fourier transform of $p(x)$,} using the Gaussian complex plane convention defined in SM Sec.~S1
\begin{equation}\lambda(\beta) = \int p(x) e^{2i\beta x} dx \end{equation} so that learning the channel is equivalent to learning $\lambda(\beta)$. 

\begin{figure*}[t]
    \centering
    \includegraphics[width=\textwidth]{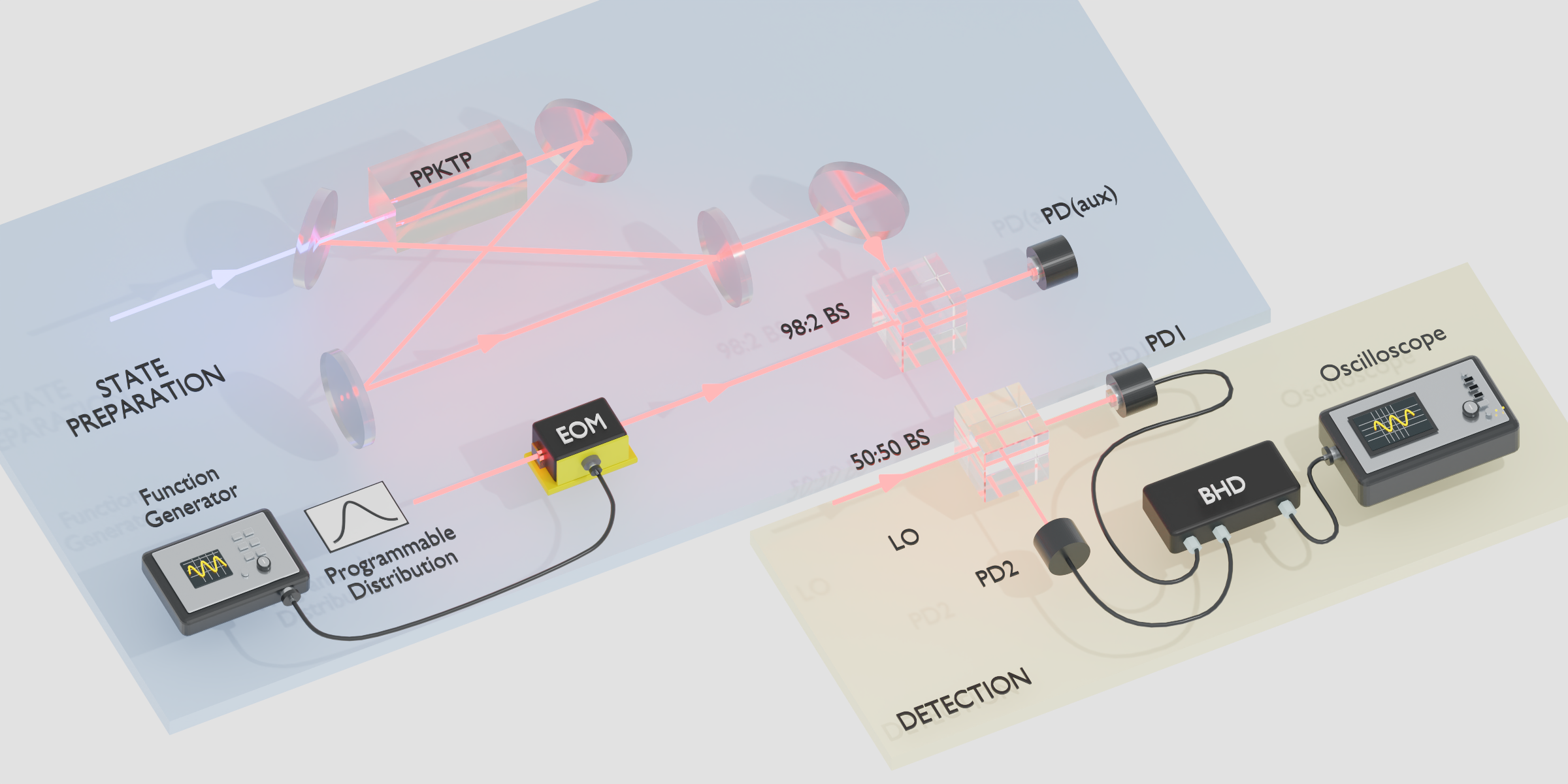}
    \caption{\textbf{Experimental \ZJ{schematic}.} \ZJ{The experiment} consists of two 
    stages: state preparation and homodyne detection. In the state 
    preparation stage, squeezed vacuum is generated via parametric 
    down-conversion in a PPKTP crystal inside a bowtie cavity (squeezer). The displacement is introduced by combining the squeezed vacuum with a \ZJ{bright} coherent beam on a 98:2 beamsplitter; 
    the coherent beam is phase-modulated by an electro-optic modulator 
    (EOM) driven by a function generator programmed with the desired 
    displacement distribution, implementing the random displacement 
    channel. An auxiliary photodetector (PD aux) monitors the displacement 
    beam power and is used to lock the displacement to the squeezing quadrature. \ZJ{To prepare a vacuum probe state, the squeezed beam is simply blocked.} In the detection stage, the displaced squeezed state is 
    interfered with a local oscillator (LO) on a 50:50 beamsplitter, and 
    the two output ports are detected by a balanced homodyne detector 
    (PD 1 and PD 2), that is locked to the 
    squeezed quadrature.}
    \label{fig:setup}
\end{figure*}

\textit{Hypothesis Testing :} We formulate a binary hypothesis testing problem \ZJ{as discriminating two channels}
\begin{flalign}
    H_0 : \lambda_0(\beta) &= e^{-\frac{1}{2}\frac{|\beta|}{\sigma_\beta}^2}\\
    H_1 : \lambda_1(\beta) &= 2 i \epsilon_0 e^{-|\beta-\beta_0|^2 / 2\sigma_\beta^2} +  e^{-\beta^2 / 2\sigma_\beta^2}  \nonumber\\ 
    & - 2i\epsilon_0 e^{-|\beta+ \beta_0|^2 / 2\sigma_\beta^2} ,\
\end{flalign}
with equal priors and $|\epsilon_0|\ZJ{\leq}1/4$ ensures that $\lambda_0(\beta)$ and $\lambda_1(\beta)$ are valid characteristic functions.

\ZJ{Note that $\lambda_0(\beta)$ is a single Gaussian centered at the origin and $\lambda_1(\beta)$} consists of three Gaussian components in Fourier space: a central peak and two symmetric side peaks at $\pm\beta_0$ of height 2$\epsilon_0$  as shown in Fig.~\ref{fig:fig1}c. \ZJ{The respective phase space distributions can be obtained by taking the Fourier transform of $\lambda_0(\beta)$ and $\lambda_1(\beta)$}
\begin{align}
    p_0(x) &= \sqrt{\frac{2}{\pi}} \sigma_\beta e^{-2\sigma_\beta^2 x^2} \\
    p_1(x) &= p_0(x) [1+4\epsilon_0 \sin{2 \beta_0 x}] .\
\end{align}
\ZJ{The resolution parameter $\beta_0$ governs the complexity of the learning task.} Increasing $\beta_0$ produces increasingly fine oscillatory structure in the phase space distribution \ZJ{$p_1(x)$,} with period $\pi/\beta_0$. Critically, $\beta_0$ does not change the Gaussian envelope or the mean-squared displacement of $p(x)$, it only redistributes probability mass over progressively finer spatial scales. Therefore, the statistical difficulty arises solely from \ZJ{resolving} information at increasingly large Fourier frequencies.

The hypothesis testing problem above is a subtask of the more general problem of reconstructing the unknown distribution $p(x)$ from homodyne measurements.

\textit{Reconstruction task:} We define reconstruction as the task of producing
an estimator $\tilde{\lambda}(\beta)$, built from $N$ homodyne
measurements, that approximates $\lambda(\beta)$ uniformly over the
bandwidth $|\beta|\leq\beta_0$ with success probability at least
$1-\delta$:
\begin{equation}
    \Pr\left[
    \sup_{|\beta|\leq\beta_0}
    |\tilde{\lambda}(\beta)-\lambda(\beta)|
    \leq\epsilon
    \right]\geq1-\delta .
    \label{eq:recon_task}
\end{equation}
\ZJ{The variable $\epsilon$ sets a tolerance for the reconstruction error $|\tilde{\lambda}(\beta)-\lambda(\beta)|$.} The worst-case pointwise sample complexity of this task is set by the highest frequency $\beta = \beta_0$. The connection between reconstruction and hypothesis testing is the following: any protocol that achieves Eq.~\eqref{eq:recon_task} can in particular evaluate $\tilde{\lambda}(\beta_0)$ and compare $\mathrm{Im}[\tilde{\lambda}(\beta_0)]$ against $\mathrm{Im}[\lambda_{0/1}(\beta_0)]$ to distinguish $H_0$ from $H_1$. Therefore the sample complexity of hypothesis testing is a lower bound on the sample complexity of reconstruction. Conversely, the upper bounds derived in the following sections for reconstruction also bound the sample complexity of hypothesis testing as a special case.

\subsection{\ZJ{The Unbiased Deconvolution Estimator}}

We now construct the estimator, $\tilde\lambda(\beta)$, whose variance exposes directly how the probe's variance along the chosen quadrature limits access to high-frequency features. The displacement channel acts on the probe state by randomly displacing it by an amount $X \sim \ZJ{p(x)}$, where \ZJ{$p(x)$} is the unknown distribution to be learned. Homodyne detection on the output state returns noisy outcomes $Y = X + Q$, where $Q \sim \mathcal{N}(0, \sigma_Q^2)$ is additive Gaussian noise arising from the probe state. For vacuum and coherent-state probes, $\sigma_Q^2 = \sigma_{\text{vac}}^2 = 1$, while for a squeezed-vacuum probe with squeezing parameter $r$ aligned with the measured quadrature, $\sigma_Q^2 = e^{-2r}$.

The measured probability distribution is the convolution
\ZJ{$p (y) = p(x) * p(q) $}, so in Fourier space the corresponding characteristic functions factorize as a product,
$\lambda_Y(\beta) = \lambda(\beta)\, \lambda_Q(\beta)$,
where $\lambda_Q(\beta) = e^{-2\sigma_Q^2|\beta|^2}$ is the characteristic function of the probe state \cite{weedbrook2012gaussian}, known {\it a priori}. The unbiased deconvolution estimator for $\lambda(\beta)$ is
\begin{equation}
\label{eq:estimator}
\tilde\lambda(\beta) = \frac{\lambda_Y(\beta)}{\lambda_Q(\beta)} ,
\end{equation}
where $\lambda_Y(\beta) = \frac{1}{N}\sum_{k=1}^N e^{2i\beta Y_k}$, is the empirical characteristic function from $N$ homodyne samples. This estimator is unbiased and since $|e^{2i\beta Y_k}| = 1$ the variance is bounded by
\begin{equation}
\label{eq:varPhiX}
\text{Var}\left(\tilde\lambda(\beta)\right) \leq \frac{1}{N|\lambda_Q(\beta)|^2} \sim \frac{1}{N}e^{4\sigma_Q^2\beta^2} .
\end{equation}
This reveals the fundamental significance of $\lambda_Q(\beta)$, which acts as an exponentially decaying bandwidth filter: the features of $\lambda$ at frequency $\beta$ are suppressed by $|\lambda_Q(\beta)|^2 = e^{-4\sigma_Q^2\beta^2}$, and recovering \ZJ{$\lambda (\beta)$ with a prescribed estimation error} requires a number of samples \ZJ{that grows exponentially with $|\beta|^2$}. Reducing the probe noise variance $\sigma_Q^2$ therefore reduces the exponent governing this sample complexity. Consequently, squeezed probes can provide an exponential reduction in the number of samples required to resolve fine-scale features of the displacement distribution, relative to vacuum or coherent probes.

\section{Results}

Here we present the experimental demonstration of the exponential learning advantage predicted in Sec.~\ref{sec:theory}. We first validate the sample-complexity separation between vacuum and squeezed probes for the binary hypothesis-testing task using Log-Likelihood Ratio (LLR) (Result~1), then extend this to the strictly harder task of reconstructing the full characteristic function $\lambda(\beta)$ over the bandwidth $|\beta|\le\beta_0$ (Result~2). In both cases we compare the experimentally measured sample complexity against the theoretical upper and lower bounds.

\subsection*{Physical Implementation}
Squeezed vacuum states are prepared at squeezing levels of  4.95\,dB, 6.20\,dB, 7.69\,dB, and 8.54\,dB via parametric down-conversion in a PPKTP crystal inside a bowtie cavity. The displacement channel is implemented by combining the squeezed vacuum with a \ZJ{bright} coherent beam on a 98:2 beamsplitter, where the coherent beam is phase-modulated by an electro-optic modulator (EOM) driven by a function generator programmed with the desired displacement distribution $p(x)$. To implement the hypothesis pair, the function generator draws displacements generated either from a single Gaussian  ($H_0$) or from oscillatory Gaussian distribution ($H_1$)  whose frequency
is set by $\beta_0$, realizing the two channels directly as stochastic displacement sequences. The homodyne detector is locked to the squeezed quadrature, and the photocurrent is demodulated at 1\,MHz to extract the displacement quadrature, as shown in Fig.~\ref{fig:setup}. The bowtie cavity design and phase locks follow the design of Ref.~\cite{shajilal202212}. Shot-noise calibration and affine correction of the displacement are applied to all data prior to analysis; across all experimental runs the displacement calibration slope remains within $s \in [0.98,\,1.04]$ and the offset satisfies $|c| \ll 1$\,SNU, confirming systematic errors are small and stable throughout. Full details of the state preparation, noise calibration, and independent sample construction are given in SM Sec~S4.

\begin{figure*}[t]
    \centering
    \includegraphics[width=0.8 \textwidth]{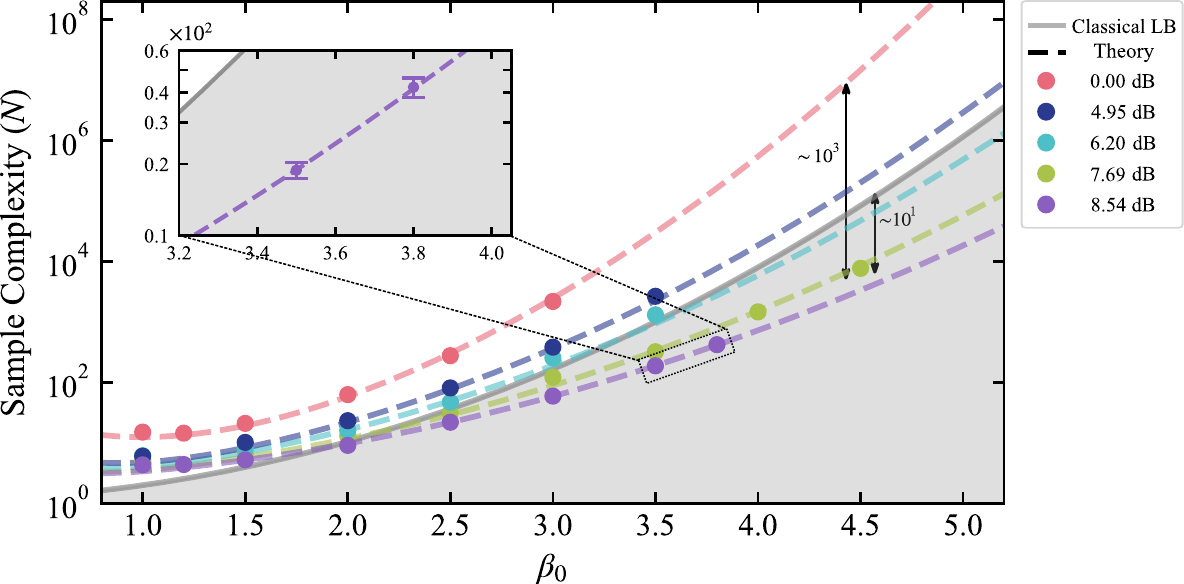}
    \caption{\textbf{\ZJ{Hypothesis tesing: exponential advantage in sample complexity.}} Minimum number of 
    homodyne measurements $N_{\min}$ required to distinguish $H_0$ from 
    $H_1$ with success probability $\geq 2/3$ using the log-likelihood 
    ratio \ZJ{(LLR)} decision rule, as a function of peak separation $\beta_0$. 
    Dashed lines show simulation results and solid markers show 
    experimental data \ZJ{using vacuum probe (red), and squeezed probe with squeezing levels of}  4.95\,dB, 
    6.2\,dB, 7.69\, dB and 8.54\,dB. The gray \ZJ{solid} line shows the 
    classical lower bound from Theorem 1. At $\beta_0 = 4.5$, the vacuum 
    probe requires $\sim 1.5 \times 10^{7}$ samples while a strongly squeezed 
    probe requires only $\sim  10^4$, a separation of three orders of 
    magnitude that grows exponentially in $\beta_0^2$. Error bars denote 
    the standard deviation over 50 independent repetitions. Parameters: 
    $\epsilon = 0.2$, $\sigma_\beta = 1$\,SNU.}
    \label{fig:llr_hypothesis}
\end{figure*}

\subsection*{Result 1: Hypothesis Testing}
The distinguishing signature between $H_0$ and $H_1$ is the emergence of Gaussian side peaks of amplitude $2\epsilon_0$ at $\pm\beta_0$ in $\lambda(\beta)$. While the deconvolution estimator \ZJ{in Eq.~\eqref{eq:estimator}} already reveals the exponential bandwidth-filtering mechanism, discrimination between two known hypotheses admits a tighter, task-optimized test. We therefore use an LLR test computed directly on the raw homodyne outcomes. Convolving each hypothesis with the probe noise preserves the Gaussian envelope of the output distribution but attaches to it a sinusoidal modulation whose visibility is exponentially attenuated by the probe noise (see SM Sec~3.1 for complete derivation). The output homodyne distributions under $H_0$ and $H_1$, denoted by $p_{0,\mathrm{out}}(x)$ and $p_{1,\mathrm{out}}(x)$, respectively, are related by:
\begin{equation}p_{1,\text{out}}(x) = p_{0,\text{out}}(x)\left[1 + 4\epsilon_0 A\sin(kx)\right] ,\end{equation}
where the fringe amplitude $A = \exp\left(-\frac{2|\beta_0|^2\sigma_Q^2}{1+4\sigma_\beta^2\sigma_Q^2}\right)$
quantifies the visibility of the distinguishing structure surviving the measurement process and $k=\frac{2\beta_0}{1+4\sigma_\beta^2\sigma_Q^2}$ is the fringe frequency. The LLR per sample is
\ZJ{$ \log[p_{\mathrm{1,out}}/ p_{\mathrm{0,out}}]= \log[1 + 4\epsilon_0 A\sin(kx)] $.}

The exponential separation between vacuum and squeezed probing arises directly from the exponential dependence of $A$ on $\sigma_Q^2$. The LLR for small $\epsilon_0$ reduces to $4\epsilon_0 A\sin(kx)$, giving an accumulated signal-to-noise ratio (SNR) that scales as $\sqrt{N}\,\epsilon_0 A$. Reliable discrimination ($\mathrm{SNR}\gtrsim 1$) therefore requires minimum samples $N_{\text{min}}$
\begin{equation}
N_{\text{min}} \propto \frac{1}{(\epsilon_0 A)^2} \propto \frac{1}{\epsilon_0^2} \exp\left(\frac{4|\beta_0|^2\sigma_Q^2}{1+4\sigma_\beta^2\sigma_Q^2}\right)
\label{eq:llr_bound}
\end{equation}
A more precise hypothesis-testing analysis, imposing $P_{\rm succ}\geq2/3$, gives the corresponding prefactor and finite fringe visibility correction ( see Supplementary Material). Since $N_\mathrm{min} \sim 1/A^2$, the required sample size grows at precisely twice the exponential rate at which $A$ decays. Squeezing reduces $\sigma_Q^2$, slowing the decay of $A$ and directly reducing the exponential cost of discrimination.

Eq.~\eqref{eq:llr_bound} demonstrates the existence of a squeezed-state protocol with exponentially lower sample complexity than a specific vacuum-probe strategy. Indeed, one might expect these results to hold for a broader class of classical optical states, namely those with positive $P$-distribution \cite{sudarshan1963equivalence}, i.e.
\begin{equation}
\rho_{\mathrm{cl}}=\int P(\alpha)\ket{\alpha}\bra{\alpha} d^2\alpha,
\qquad P(\alpha)\geq 0.
\end{equation}
However, it does not rule out the possibility that some other vacuum state protocol using a different estimator, adaptive measurements, or arbitrary post-processing  could match the squeezed probe's performance. We now close this gap with an information-theoretic lower bound obtained by bounding the success probability of discriminating $H_0$ from $H_1$ via the distinguishability of the corresponding output quantum states. Since $H_0$ and $H_1$ differ at $\beta_0$ by $|\lambda_0(\beta_0)-\lambda_1(\beta_0)|=\Delta_{\beta_0}$, any protocol that reconstructs $\lambda(\beta)$ with error less than $\Delta_{\beta_0}/2$ over $|\beta|\le\beta_0$ can in particular discriminate the two hypotheses; the resulting lower bound on discrimination therefore also lower-bounds the more general reconstruction task of Result~2.

\begin{theorem}
Let $p(x)$ be a probability distribution of displacements along the $x$ direction with characteristic function $\lambda(\beta)$. Any protocol using classical probe states (nonnegative joint Glauber–Sudarshan \(P\) representation) that is guaranteed for every $p(x)$ to determine $\lambda(\beta)$ to within error $\epsilon$, satisfying $
0<\epsilon<
\epsilon_0\left(1-e^{-2|\beta_0|^2/\sigma_\beta^2}\right)$, for all $\beta$ with $|\beta| \leq \beta_0$, with success probability greater than $1-\delta$, with \(0<\delta<1/2\),  requires a number of parallel channel uses $N$ such that
$$
N\geq
\frac{\log[1+4(1-2\delta)^2]}{8\epsilon_0^2}
\exp\!\left[
\frac{|\beta_0|^2}{\sigma_\beta^2}
\left(
1-\frac{1}{\sqrt{1+4\sigma_\beta^2}}
\right)
\right].
$$
For fixed squeezed-vacuum inputs with measured quadrature variance \(a=e^{-2r}\), the same bound holds with \(1+4\sigma_\beta^2\) replaced by \(1+4a\sigma_\beta^2\).

\end{theorem}

% \begin{theorem}
% Let $p(x)$ be a probability distribution of displacements along the $x$ direction with characteristic function $\lambda(\beta)$. Using any classical-state probe to determine $\lambda(\beta)$ to within error $\epsilon$ for all $\beta$ with $|\beta| \leq \beta_0$, with success probability greater than $1-\delta$, requires a number of channel uses $N$ such that
% \begin{equation}
% N\geq
% \frac{\log(13/9)}{8\epsilon_0^2}
% \exp \left[
% \frac{|\beta_0|^2}{\sigma_\beta^2}
% \left(
% 1-\frac{1}{\sqrt{1+4a\sigma_\beta^2}}
% \right)
% \right]
% \end{equation}
% where $a = \sigma_Q^2 = 1$ for vacuum. The analogous bound for a squeezed-vacuum probe has $a = e^{-2r}$.\\
% \end{theorem}

The upper and lower bounds share the same exponential structure in $\beta_0^2$: the separation is a fundamental information-theoretic property of the problem, not an \ZJ{artefact} of the estimator or decision rule. Any protocol regardless of estimator, adaptive strategy, or post-processing  could not overcome this barrier. The full proof is given in SM Sec.~S2.

Eq.~\eqref{eq:llr_bound} shows that for any fixed squeezing level $r$, the sample complexity required to successfully discriminate the two hypothesis diverges exponentially in $\beta_0^2$. However, when combined with Theorem~1, for $\sigma_Q^2 < \frac{\sqrt{1+4\sigma_\beta^2}-1}{4\sigma_\beta^2}$, the sample complexity of the squeezed-probe protocol asymptotically falls exponentially far below the classical lower bound. Using Eq.~\eqref{eq:varPhiX} a similar result follows for the full reconstruction task. We now present the following corollary that gives the conditions on the squeezing level that are required for the number of samples to remain constant.

\begin{figure*}[t]
    \centering
    \includegraphics[width=0.8\textwidth]{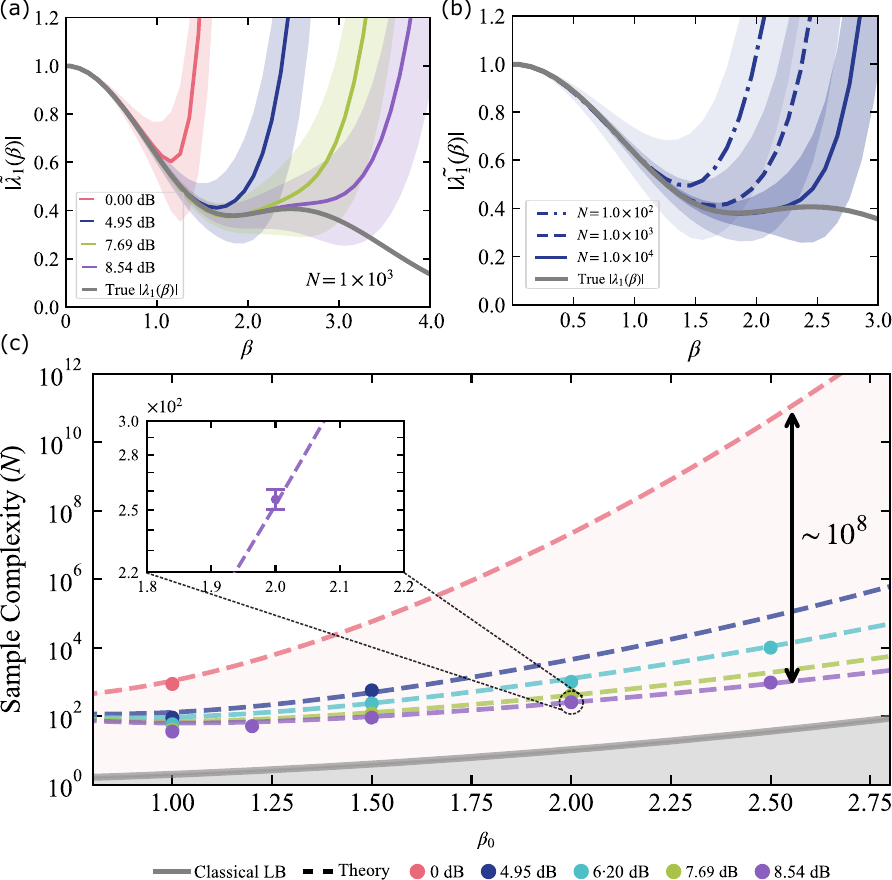}
    \caption{\textbf{\ZJ{Full-bandwidth reconstruction of characteristic functions.}}
    (a) Experimental characteristic function reconstruction $\lambda(\beta)$ whose highest frequency component $\beta_0 = 2.5$ for a fixed number of homodyne samples to be $N = 1 \times 10^3$ but using probes of different variances. The solid lines represent the average outcome of 2000 runs of reconstruction task as a function of number of homodyne samples and the shadings are 1$\sigma$ standard deviation.
    (b) The same reconstruction task as given in panel (a) at $\beta_0 = 2.5$ but for a given probe of variance -4.95 dB and varying the number of homodyne samples $(N)$  (c) Minimum number of homodyne measurements $N_{\min}$ required to estimate 
    $\lambda(\beta)$ with success probability $\geq 2/3$ as a function 
    of the \ZJ{highest Fourier frequency} $\beta_0$. Dashed lines show the theoretical $N_{\text{min}}$ and solid markers show experimental data at the 
    corresponding squeezing levels. The separation between vacuum and 
    squeezed probes grows exponentially in $\beta_0^2$, reaching 
    $\sim$8 orders of magnitude within the experimentally accessible 
    range. The vacuum probe data is limited to $\beta_0 \leq 1.0$ as 
    the required sample size exceeds the available dataset beyond this 
    point. Parameters: $\epsilon = 0.2$, $\sigma_\beta = 1$\,SNU, 
    target success probability $2/3$.}
    \label{fig:reconstruction}
\end{figure*}

\begin{corollary}
With $r = \ln(2\beta_0)$,
the exponent in Eq.~\eqref{eq:varPhiX} (also Eq.~\eqref{eq:llr_bound} asymptotically) is $1$ as $\beta_0\to\infty$ for any fixed $\sigma_\beta$, giving a $\beta_0$-independent upper bound
\begin{equation}
N =O(1)\;.
\end{equation}
\end{corollary}

A squeezing level that grows only logarithmically with the target resolution, $r \sim \ln2\beta_0$, is therefore sufficient to keep the sample complexity of this single-mode task bounded by a constant, even as $\beta_0\to\infty$ making the underlying discrimination task arbitrarily hard. This constant sample complexity regime is the single-mode, entanglement-free analog of the result of Ref.~\cite{oh2024entanglement}, who achieve $O(1)$ sample complexity in the multi-mode setting only through EPR-entangled probes.

To validate these predictions experimentally, we implement the LLR decision rule directly on homodyne data. For fixed squeezing levels and $\beta_0$, the LLR is evaluated as $\mathrm{LLR} = \sum_{k=1}^N \log(1 + 4\epsilon_0 A_\mathrm{sq}\sin(k_\mathrm{sq} x_k))$, and $H_1$ is declared if $\mathrm{LLR} > 0$. For each candidate $N$, the success probability is estimated from 2000 independent hypothesis-testing realizations as the fraction of correct decisions, and the minimum sample complexity $N_\mathrm{min}$ is determined by binary search as the smallest $N$ achieving $P_\mathrm{succ} \geq 2/3$. The complete binary search is repeated 50 times using independent random subsampling within the experimental dataset ensemble to characterize the Monte Carlo variation of $N_\mathrm{min}$.

The results are shown in Fig.~\ref{fig:llr_hypothesis} alongside theoretical prediction \ZJ{(dashed curves)} from Eq.~\eqref{eq:llr_bound} and the classical lower bound from Theorem~1 \ZJ{(solid grey curve)}. The experimental data tracks the simulation closely across all squeezing levels, confirming that the LLR decision rule performs as predicted by the theoretical framework. The exponential separation between vacuum and squeezed probing is visible across the full range of $\beta_0$ and grows with it: at $\beta_0 = 4.5$, for instance, the vacuum probe requires ${\sim}10^7$ samples while the 7.69\,dB squeezed probe requires only ${\sim} 10^4$ — already a separation of three orders of magnitude, which widens further as $\beta_0$ increases. \ZJ{More remarkably, we demonstrate a quantum learning advantage enabled by squeezed light probes, requiring ten times fewer samples than classical methods—i.e. classical probes combined with arbitrary measurements, adaptive operations and data processing.}

\subsection*{Result 2: Distribution Reconstruction}

\ZJ{Recall the reconstruction task} in Eq.~\eqref{eq:recon_task}, the full-bandwidth reconstruction of $\lambda(\beta)$ requires faithful estimation of $\lambda(\beta)$ over the entire bandwidth $|\beta| \leq \beta_0$, a strictly harder task than hypothesis testing. Since the deconvolution estimator's variance (Eq.~\eqref{eq:varPhiX}) grows with $|\beta|$, the worst-case sample complexity over this bandwidth is set by its highest frequency, $\beta = \beta_0$. 
%Any sample size sufficient to resolve $\lambda(\beta_0)$ to absolute error $\varepsilon$ is therefore also sufficient at every other frequency in the band. 

Applying Hoeffding's inequality to the deconvolution estimator $\tilde\lambda(\beta_0)$ (see SM Sec~3.2) provides a sufficient condition for the estimation error probability to be
at most $\delta$ \cite{hoeffding1963probability}:
\begin{equation}
    N \geq \frac{2\log(2/\delta)}{\epsilon^2}\,e^{4\sigma_Q^2\beta_0^2} .\
    \label{eq:hoeffding_bound}
\end{equation}
\ZJ{Here,} we set the reconstruction error $\epsilon $ to be slightly below $ \epsilon_0(1 - e^{-2|\beta_0|^2/\sigma_\beta^2})$ (i.e. just under half of the amplitude of the distinguishing feature at $\beta_0$), and $\delta = 1/3$: \ZJ{the error probability equals $1-P_{\mathrm{succ}}$.} The Hoeffding bound is a worst-case upper bound. The actual minimum $N$ lies below it by a constant factor, confirmed numerically to be independent of $\beta_0$ and $\sigma_Q^2$, leaving the exponential scaling exact \ZJ{(see SM)}. Since hypothesis testing is a special case of reconstruction, the lower bound of Theorem~1 applies directly. 

To validate these predictions experimentally, we implement the \ZJ{full-bandwidth} reconstruction \ZJ{learning task} on a finite frequency grid \ZJ{$\mathcal{G}=\{\beta_1,\ldots,\beta_{n_\beta-1}, \beta_0\}\subset[0,\beta_0]$}, \ZJ{discritising the full bandwidth into} finite-resolution operational \ZJ{regimes.} The bandwidth edge $\beta_0$ is included in $\mathcal{G}$ and the $n_\beta - 1$ frequencies are drawn uniformly at 
random from $[0, \beta_0]$. For each trial with $N$ homodyne samples, we 
estimate the characteristic function simultaneously 
at \ZJ{all $\beta \in \mathcal{G}$}. At each frequency $\beta$, the 
deconvolution estimator gives $\tilde{\lambda}(\beta) = e^{2\sigma_{Q,\mathrm{eff}}^2\beta^2}\frac{1}{N}\sum_{k=1}^{N} \ZJ{e^{2i\beta y_k} }$, \ZJ{using Eq.~\eqref{eq:estimator}, where $\{y_k \}$ denotes the homodyne measurement outcomes}. Reconstruction is declared successful when $|\tilde{\lambda}(\beta)-\lambda_1(\beta)| \leq\epsilon \text{ for all }\beta\in\mathcal{G}.$ For each candidate N, the success probability is estimated over 2000 realizations, and the minimum N achieving a success probability $\geq2/3$ is determined by binary search and repeated 50 times to characterize its statistical variability. For each squeezing level (including the vacuum probe) this was repeated for larger $\beta_0$ values until the sample complexity exceeds the available dataset size.
%beyond this point, a direct practical consequence of the exponential scaling that makes vacuum reconstruction infeasible at moderate $\beta_0$ while squeezed probing remains tractable. 

%The results are shown in Fig.~\ref{fig:reconstruction} alongside the theoretical lower bound curves from Theorem~1. The experimental $N_{\min}$ values track the theoretical scaling (Eq.~\eqref{eq:hoeffding_bound}) closely across all squeezing levels, confirming the exponential growth in $\beta_0^2$ predicted by the reconstruction bound. The exponential separation between squeezing levels is clearly visible and grows with $\beta_0$. At $\beta_0 = 3.8$, the 8.5\,dB squeezed probe requires approximately 18 orders of magnitude fewer samples than the vacuum probe with homodyne detection. The vacuum probe data is limited to $\beta_0 \leq 1.5$ because the sample complexity exceeds the available dataset size beyond this point, a direct practical consequence of the exponential scaling that makes vacuum reconstruction infeasible at moderate $\beta_0$ while squeezed probing remains tractable. The consistent gap across all squeezing levels confirms that it reflects the sub-optimality of the specific estimator relative to the optimal Neyman--Pearson test, rather than any experimental imperfection. 

The results are shown in Fig.~\ref{fig:reconstruction} alongside the theoretical lower bound curves from Theorem~1. Panels (a) and (b) illustrate the reconstruction directly. Panel (a) shows the reconstructed $|\tilde{\lambda}_1(\beta)|$ for $\beta_0 = 2.5$ as a function of $\beta$ for fixed $N = 1\times10^3$ samples across different squeezing levels, with shading indicating $1\sigma$ standard deviation over 2000 reconstruction runs. The vacuum probe (0 dB) reconstruction becomes unreliable beyond $\beta \approx 1$. The reconstructed curve deviates dramatically from the true $|\lambda_1(\beta)|$ (gray) and the uncertainty band grows rapidly. Stronger squeezing faithfully reconstructs $\lambda_1(\beta)$ to progressively higher frequencies with the same number of samples, with the 8.54 dB probe tracking the true characteristic function well beyond $\beta = 2$. Panel (b) shows the same reconstruction for a fixed squeezing level of 4.95 dB at different sample sizes $N \in \{10^2, 10^3, 10^4\}$. At small $N$ the reconstruction is noisy and the high-frequency features are unresolved. As $N$ increases the estimate converges to the true $|\lambda_1(\beta)|$, with the characteristic dip and recovery of the three-peak structure becoming clearly resolved at $N = 10^4$. The experimental minimum sample complexity 
$N_\mathrm{min}$ for the reconstruction task \ZJ{closely} track the theoretical scaling (Eq.~\eqref{eq:hoeffding_bound}) across all squeezing levels, confirming the exponential growth in $\beta_0^2$ predicted by the reconstruction bound as shown in panel c. The exponential separation between \ZJ{squeezed and vacuum probes} is clearly visible and grows with $\beta_0$. At $\beta_0 = 2.5$, the vacuum probe requires ${\sim} 10^8$ times more samples than the 8.54 dB squeezed probe to achieve successful reconstruction over $|\beta| \leq \beta_0$, directly demonstrating the exponential quantum advantage in the 
reconstruction task. 
%The vacuum probe data is limited to $\beta_0 \leq 1.0$ because the sample complexity exceeds the available dataset size beyond this point, a direct practical consequence of the exponential scaling that makes vacuum reconstruction infeasible at moderate $\beta_0$ while squeezed probing remains tractable. 
%The consistent gap across all squeezing levels confirms that it reflects the sub-optimality of the specific estimator relative to the optimal Neyman--Pearson test, rather than any experimental imperfection. 

\section{Conclusion \& Discussion}

We have demonstrated an exponential quantum learning advantage using only a single squeezed mode, without entanglement or an increase in system size. For a single-mode, single-quadrature random displacement channel, we have proven an information-theoretic lower bound establishing that any vacuum-state protocol requires a number of channel uses exponential in the target resolution $\beta_0$, and we have shown experimentally that a squeezed-vacuum probe accomplishes the same task with exponentially fewer samples. Crucially, this separation is tunable at will: for fixed squeezing above the critical threshold, the sample complexity eventually falls below the vacuum lower bound as the target resolution is increased (Eq.~\eqref{eq:llr_bound}). Instead, by letting the squeezing grow only logarithmically with the task's difficulty, $r\sim\ln2\beta_0$, is enough to keep the sample complexity bounded by a constant even as $\beta_0$ grows asymptotically large (Corollary~1). In both regimes, the resource being tuned is not entanglement or system size but the accessible bandwidth of a single mode, set entirely by the probe's variance. These results raise a broader question about which resources are fundamentally responsible for such advantages, and open the door to identifying accessible bandwidth, rather than entanglement or system size alone, as an independent resource for scalable quantum learning.

A natural question is whether the single-mode exponential advantage follows trivially from the multi-mode results of Oh et al. by setting $n=1$. It does not. In the $n$-mode displacement channel learning problem, the exponential separation is parametric in $n$ where entanglement-free strategies require $\exp(\Omega(n))$ samples while EPR-assisted strategies require $O(1)$ samples independent of $n$. At $n=1$, this separation collapses and the number of parameters required to characterize the single-mode channel is $\sim k^2$ where $k$ is the number of Fourier components. The exponential separation we establish is of a qualitatively different character: it is parametric in the target resolution $\beta_0$ not in the system dimension, and arises from the exponential suppression of high-frequency characteristic function components by the probe's noise variance. The single-mode setting is therefore not a degenerate case of the multi-mode problem but a distinct and independently motivated learning task, whose difficulty is controlled by a tunable resolution parameter rather than by system dimensionality.

The exponential separation established above has a transparent physical origin. The probe's characteristic function $|\lambda_Q(\beta_0)| = e^{-2\sigma_Q^2\beta_0^2}$ decays exponentially in $\beta_0^2$, acting as a low-pass filter that suppresses access to the high-frequency features that distinguish the two hypotheses. Squeezing reduces $\sigma_Q^2$ from $\sigma_\text{vac}^2 = 1$ to $\sigma_{Q,\text{eff}}^2 \ll 1$, exponentially slowing this decay and extending the accessible bandwidth. The advantage is therefore metrological in origin. It arises from variance control in Fourier space rather than from parameter estimation precision in the usual sense  and places this problem well within the framework of statistical learning theory.

\ZJ{Importantly, in contrast to prior quantum learning results where the exponential separation is parametric in the number of modes $n$, here,} the system dimension is fixed, the measurement apparatus is fixed, and the resource comparison is unambiguous. The only difference between the vacuum and squeezed protocols is the probe state. This makes the exponential advantage of squeezing a clean, resource-accountable statement and demonstrates that entanglement is not a necessary condition for exponential quantum learning advantages when the task demands resolution of fine structure in phase space.

\section*{Note added}
During the finalization of this manuscript, we became aware of related work by Kannan et al.~\cite{kannan2026exponential} that establishes exponential learning advantages for several sensing architectures, including squeezed Gaussian probes compared with classical coherent probes. Our work provides a complementary direct optical implementation using squeezed-vacuum preparation and homodyne detection for one-quadrature displacement distributions, together with fixed-squeezing benchmarks against an arbitrary-measurement classical lower bound. Both of these results provides further 
evidence that the necessity of entanglement in prior quantum learning results is a consequence of the specific problem structure rather than a fundamental requirement of quantum learning. 

\section*{Author Contributions}

\noindent A.A.B. and L.O.C. conceived and developed the theoretical framework. A.A.B., J.Q, J.Z and B.S conceived the experiment design. A.A.B. performed the experiment and data analysis. A.D, O.E., L.G., S.A. reviewed the calculations and contributed to the writing of the manuscript. All authors discussed the results and contributed to the final manuscript.

\section*{Data \& Code Availability}

\noindent Correspondence and requests for data and materials should be addressed to A.A.B. , L.O.C. and J.Z.

\begin{acknowledgments}
This project is supported by the National Research Foundation of Korea (RS-2024-00509800). J. Zhao acknowledges support from the Australian Research Council DECRA Fellowship (Grand No. DE260101046). This work is also supported by the National Research Foundation, Singapore through the National Quantum Office, hosted in A*STAR, under its Centre for Quantum Technologies Funding Initiative (S24Q2d0009).

L.O.C.~was supported in part by
the National Science Foundation under Award No. 2533041 (NQVL:Design:ORAQL), ONR MURI, NSF QLCI (award No.~OMA-2120757), and the NSF STAQ program. 

\end{acknowledgments}
\section*{Competing Interests}
The authors declare that they have no competing financial interests.

\newpage

\bibliography{apssamp}% Produces the bibliography via BibTeX.

\clearpage
\onecolumngrid

\setlength{\textwidth}{16cm}
\setlength{\oddsidemargin}{0cm}
\setlength{\evensidemargin}{0cm}

\begin{center}
{\Large\bfseries Supplementary Material}
\end{center}

\vspace{1em}

\setcounter{section}{0}
\setcounter{equation}{0}
\setcounter{figure}{0}
\setcounter{table}{0}

%\startcontents[supp]
%\printcontents[supp]{}{1}{}
%\newpage

\counterwithout{equation}{section}

\renewcommand{\thesection}{S\arabic{section}}
\renewcommand{\thefigure}{S\arabic{figure}}
\renewcommand{\thetable}{S\arabic{table}}
\renewcommand{\theequation}{S\arabic{equation}}

%\setcounter{tocdepth}{2}
%\startcontents[supp]
%\printcontents[supp]{}{1}{}
%\tableofcontents
%\newpage
\makeatletter
\providecommand{\@pnumwidth}{1.55em}
\providecommand{\@tocrmarg}{2.55em}
\providecommand{\@dotsep}{4.5}
\renewcommand*\l@section{\@dottedtocline{1}{0em}{2.6em}}
\renewcommand*\l@subsection{\@dottedtocline{2}{2.6em}{2.6em}}
\renewcommand*\l@subsubsection{\@dottedtocline{3}{5.2em}{3.6em}}
\@ifundefined{c@tocdepth}{\newcounter{tocdepth}}{}
\makeatother
\setcounter{tocdepth}{3}
\addtocontents{toc}{\protect\supptocstart}
{\makeatletter
\let\supp@contentsline\contentsline
\def\contentsline#1#2#3#4{}
\def\supptocstart{\let\contentsline\supp@contentsline}
\tableofcontents
\makeatother}
\newpage
\section{Notation and Problem Setup}
\label{sec:notation}

This section establishes the notation and operator 
conventions used throughout this theoretical and experimental work, 
introduces the displacement channel and the binary 
hypothesis pair, and derives the characteristic 
functions corresponding to the experimental 
distributions. All subsequent proofs follow this 
setup.

\subsection{Operator conventions}

We use the creation and annihilation operators 
$a^\dagger$ and $a$ satisfying $[a,a^\dagger]=1$, 
and define the quadrature operators
\begin{equation}
    \hat{x} = a + a^\dagger, 
    \qquad 
    \hat{p} = -i(a - a^\dagger),
    \qquad 
    [\hat{x},\hat{p}] = 2i.
\end{equation}
This convention normalizes the vacuum noise to unity: 
$\langle 0|\hat{x}^2|0\rangle = 
\langle 0|\hat{p}^2|0\rangle = 1$, matching the 
shot-noise unit (SNU) calibration used in the 
experiment. The displacement operator along $\hat{x}$ 
by amount $x \in \mathbb{R}$ is
\begin{equation}
    U_x = D(x/2) = e^{-ix\hat{p}/2},
    \qquad 
    U_x|0\rangle = |x\rangle,
\end{equation}
where $|x\rangle$ is the coherent state satisfying 
$\langle\hat{x}\rangle = x$ and 
$\langle\hat{x}^2\rangle = 1 + x^2$. We also define
\begin{equation}
    W_\beta := e^{i\beta\hat{x}} = D(i\beta),
\end{equation}
which satisfies $U_y^\dagger W_\beta U_y = 
e^{i\beta y}W_\beta$ since 
$U_y^\dagger\hat{x}U_y = \hat{x} + y$.

\subsection{Displacement channel and characteristic 
function}

The random displacement channel implemented in the 
experiment is
\begin{equation}
    \Lambda(\rho) = \int dx\, p(x)\, 
    U_x \rho U_x^\dagger,
    \label{eq:channel}
\end{equation}
where $p(x)$ is the unknown classical displacement 
distribution along $\hat{x}$. The Heisenberg-picture action  $\mathcal{E}^\dagger$ of this channel on $W_\beta$ gives \cite{weedbrook2012gaussian}
\begin{equation}
    \mathcal{E}_j^\dagger(W_\beta) 
    = \lambda_j(\beta) W_\beta,
    \qquad 
    \lambda_j(\beta) = \int_{\mathbb{R}} 
    p_j(y)\, e^{i\beta y}\, dy,
    \label{eq:charfun_def}
\end{equation}
which is the restriction of the symplectic Fourier 
relation for a random-displacement channel to noise 
supported on a single quadrature. The characteristic 
function $\lambda_j(\beta)$ fully characterizes the 
channel, so learning the channel is equivalent to 
learning $\lambda_j(\beta)$.

We work with the rescaled Fourier coordinate 
$\beta' = \beta/2$, under which the estimator 
becomes
\begin{equation}
    \lambda_j'(\beta') = \lambda_j(2\beta') 
    = \int_{\mathbb{R}} p_j(y)\, 
    e^{2i\beta' y}\, dy,
    \label{eq:rescaled_estimator}
\end{equation}
consistent with the empirical estimator 
$\hat{\lambda}'(\beta') = 1/N\sum_k e^{2i\beta' x_k}$ 
used in the main text. In these rescaled coordinates 
the learning task is to reconstruct $\lambda'(\beta')$ 
for $|\beta'| \leq \beta_0$, rather than 
$\lambda(\beta)$ for $|\beta| \leq 2\beta_0$. All 
subsequent analysis uses the rescaled convention and 
we drop the prime notation for clarity, writing 
$\lambda(\beta) \equiv \lambda'(\beta')$ and 
$\beta \equiv \beta'$ throughout.

\subsection{Binary Hypothesis}

We now introduce the specific distributions used to 
establish the sample-complexity lower bound. The 
experimental displacement distributions are
\begin{align}
    H_0 & : p_0(x) = \sqrt{\frac{2}{\pi}}\,\sigma_\beta\, 
    e^{-2\sigma_\beta^2 x^2},
    \label{eq:p0}\\
    H_1 &: p_1(x) = p_0(x)
    \left[1 + 4\epsilon_0 \sin(2|\beta_0| x)\right],
    \label{eq:p1}
\end{align}
with $|\epsilon_0| \leq 1/4$ ensuring $p_1(x) \geq 0$. 
The corresponding characteristic functions, computed 
from Eq.~\eqref{eq:charfun_def} under the rescaled 
convention, are:
\begin{align}
    \lambda_0(\beta) 
    &= \int \sqrt{\frac{2}{\pi}}\,\sigma_\beta\, 
    e^{-2\sigma_\beta^2 x^2} e^{2i\beta x}\,dx 
    = e^{-\beta^2/2\sigma_\beta^2},
    \label{eq:lambda0}\\
    \lambda_1(\beta) 
    &= \int \sqrt{\frac{2}{\pi}}\,\sigma_\beta\, 
    e^{-2\sigma_\beta^2 x^2}
    \left[1 + 4\epsilon_0\sin(2\beta_0x)\right]
    e^{2i\beta x}\,dx \notag\\
    &= e^{-\beta^2/2\sigma_\beta^2} 
    + 2i\epsilon_0\, 
    e^{-(\beta-\beta_0)^2/2\sigma_\beta^2}
    - 2i\epsilon_0\, 
    e^{-(\beta+\beta_0)^2/2\sigma_\beta^2}.
    \label{eq:lambda1}
\end{align}
The two hypotheses share the same Gaussian envelope and second moment, 
they differ only in the presence of side peaks at $\pm\beta_0$ in 
characteristic function space, corresponding to 
fine oscillatory structure of period $\pi/\beta_0$ 
in $p_1(x)$. The statistical difficulty of 
distinguishing them increases with $\beta_0$ 
because the side peaks move to higher Fourier 
frequencies where the probe's characteristic 
function is more strongly attenuated.

The binary hypothesis test is formulated with 
equal priors. A uniformly random binary sign 
$s \in \{-1,+1\}$ determines which hypothesis 
generated the data. Any lower bound 
on the sample complexity of this hypothesis test 
is also a lower bound on the sample complexity of reconstruction, since reconstructing $p(x)$ to sufficient accuracy necessarily allows the two hypotheses to be distinguished. We now derive this lower bound.

\section{Information theoretic Lower Bound on Sample Complexity}
\label{sec:lower_bound}

We now derive a lower bound on the number of channel uses required
to distinguish the two hypotheses introduced in Sec.~\ref{sec:notation}.
We first consider a fixed squeezed-vacuum probe and allow the receiver
to perform an arbitrary measurement on the resulting outputs, including
collective measurements across all channel uses. This establishes a
measurement-independent lower bound for squeezed-vacuum probing.

%\subsection{Lower bound for arbitrary measurements with a fixed squeezed-vacuum probe}

We take the probe state to be a single-mode squeezed vacuum
$\ket{0_r}$, with squeezing applied along the displaced quadrature.
It is convenient to denote its quadrature variance by
\begin{equation}
V_r:=\operatorname{Var}_{\ket{0_r}}(\hat{x})=e^{-2r}.
\label{eq:probe-variance}
\end{equation}
Throughout this section we use the canonical quadrature
\begin{equation}
\hat K:=\frac{\hat p}{2},
\qquad
[\hat x,\hat K]=i,
\label{eq:arb-K-def}
\end{equation}
which follows from the convention $[\hat x,\hat p]=2i$.
Thus, $(\hat x,\hat K)$ form a canonically normalized quadrature pair.
The conjugate-quadrature variance is therefore
\begin{equation}
\operatorname{Var}_{\ket{0_r}}(\hat K)
=
\operatorname{Var}_{\ket{0_r}}(\hat p/2)
=
\frac{1}{4V_r}.
\label{eq:probe-conjugate-variance}
\end{equation}

For the two displacement distributions defined in
Eqs.~\eqref{eq:p0}–\eqref{eq:p1}, the variance of the Gaussian
displacement distribution $p_0$ is
\begin{equation}
S:=\operatorname{Var}{p_0}(X)
=\frac{1}{4\sigma\beta^2},
\qquad
\omega:=2|\beta_0|.
\label{eq:arb-S-omega}
\end{equation}
Here, $S$ characterizes the classical displacement noise, while
$\omega$ is the spatial frequency of the oscillatory modulation
in $p_1(x)$.

Under the null hypothesis, the output quadrature is the sum of
the input quantum quadrature and the random displacement, $\hat{x}_{{\mathrm{out}}}
=
\hat{x}_{{\mathrm{in}}}+X.$
Since the squeezed-vacuum fluctuations and the classical displacement
are independent, their variances add. Hence the total output variance is
\begin{equation}
V:=\operatorname{Var}{H_0}(\hat{x}_{{\mathrm{out}}})
=V_r+S.
\label{eq:total-output-variance}
\end{equation}

The corresponding single-use output states are
\begin{equation}
\rho_j
:=
\int_{\mathbb R}p_j(x),
U_x\ket{0_r}\bra{0_r}U_x^\dagger,dx,
\qquad j\in{0,1}.
\label{eq:arb-output-states}
\end{equation}

The following theorem shows that the distinguishability of these two
output states remains exponentially suppressed at large $\beta_0$,
even when the receiver is allowed an arbitrary collective measurement.

\section{Information theoretic Lower Bound on Sample Complexity}
\label{sec:lower_bound}

We now derive a lower bound on the number of channel uses required
to distinguish the two hypotheses introduced in Sec.~\ref{sec:notation}.
We first consider a fixed squeezed-vacuum probe and allow the receiver
to perform an arbitrary measurement on the resulting outputs, including
collective measurements across all channel uses. This establishes a
measurement-independent lower bound for squeezed-vacuum probing.

%\subsection{Lower bound for arbitrary measurements with a fixed squeezed-vacuum probe}

We take the probe state to be a single-mode squeezed vacuum
$\ket{0_r}$, with squeezing applied along the displaced quadrature.
It is convenient to denote its quadrature variance by
\begin{equation}
V_r:=\operatorname{Var}_{\ket{0_r}}(\hat{x})=e^{-2r}.
\label{eq:probe-variance}
\end{equation}
Throughout this section we use the canonical quadrature
\begin{equation}
\hat K:=\frac{\hat p}{2},
\qquad
[\hat x,\hat K]=i,
\label{eq:arb-K-def}
\end{equation}
which follows from the convention $[\hat x,\hat p]=2i$.
Thus, $(\hat x,\hat K)$ form a canonically normalized quadrature pair.
The conjugate-quadrature variance is therefore
\begin{equation}
\operatorname{Var}_{\ket{0_r}}(\hat K)
=
\operatorname{Var}_{\ket{0_r}}(\hat p/2)
=
\frac{1}{4V_r}.
\label{eq:probe-conjugate-variance}
\end{equation}

For the two displacement distributions defined in
Eqs.~\eqref{eq:p0}–\eqref{eq:p1}, the variance of the Gaussian
displacement distribution $p_0$ is
\begin{equation}
S:=\operatorname{Var}{p_0}(X)
=\frac{1}{4\sigma\beta^2},
\qquad
\omega:=2|\beta_0|.
\label{eq:arb-S-omega}
\end{equation}
Here, $S$ characterizes the classical displacement noise, while
$\omega$ is the spatial frequency of the oscillatory modulation
in $p_1(x)$.

Under the null hypothesis, the output quadrature is the sum of
the input quantum quadrature and the random displacement, $\hat{x}_{{\mathrm{out}}}
=
\hat{x}_{{\mathrm{in}}}+X.$
Since the squeezed-vacuum fluctuations and the classical displacement
are independent, their variances add. Hence the total output variance is
\begin{equation}
V:=\operatorname{Var}{H_0}(\hat{x}_{{\mathrm{out}}})
=V_r+S.
\label{eq:total-output-variance}
\end{equation}

The corresponding single-use output states are
\begin{equation}
\rho_j
:=
\int_{\mathbb R}p_j(x),
U_x\ket{0_r}\bra{0_r}U_x^\dagger,dx,
\qquad j\in{0,1}.
\label{eq:arb-output-states}
\end{equation}

The following theorem shows that the distinguishability of these two
output states remains exponentially suppressed at large $\beta_0$,
even when the receiver is allowed an arbitrary collective measurement.

\paragraph{Theorem 1 (arbitrary output measurements; fixed squeezed-vacuum
input). \label{thm:arb-measurement-lower-bound}}
Suppose that the unknown channel is chosen with equal prior probability
from the two hypotheses in Eq.~\eqref{eq:p0}–\eqref{eq:p1}. On every
channel use the input is the fixed state $\ket{0_r}$. Thus, after $N$
uses, the two possible output states are
$\rho_0^{\otimes N}$ and $\rho_1^{\otimes N}$.
The receiver may perform any POVM on the $N$ outputs, including a
collective POVM with an arbitrary hypothesis-independent ancilla.
If the receiver succeeds with probability at least $2/3$, then
\begin{equation}
N\geq
\frac{\log(13/9)}{8\epsilon_0^2}
\exp \left[
\frac{|\beta_0|^2}{\sigma_\beta^2}
\left(
1-\frac{1}{\sqrt{1+4V_r\sigma_\beta^2}}
\right)
\right].
\label{eq:arb-simple-lower-bound}
\end{equation}

\paragraph{Proof.}
We first derive a measurement-independent upper bound on the
distinguishability of the two single-use output states. The argument
has three ingredients: (i) an exact evaluation of an order-two
quantum divergence between $\rho_1$ and $\rho_0$, (ii) the
data-processing inequality under an arbitrary measurement, and
(iii) the relation between order-two divergence and binary
hypothesis-testing success probability. The general information-theoretic
relations used in the last two steps are collected in
Appendix~\ref{app:renyi-discrimination}.

We work in the generalized eigenbasis of the canonical quadrature
\begin{equation}
    \hat K:=\frac{\hat p}{2},
    \qquad
    [\hat x,\hat K]=i,
\end{equation}
so that
\begin{equation}
    \hat K\ket{k}=k\ket{k}.
\end{equation}
The use of $\hat K=\hat p/2$ is therefore simply a consequence of the
quadrature convention $[\hat x,\hat p]=2i$: the pair $(\hat x,\hat K)$
obeys the standard canonical commutation relation.

The squeezed-vacuum wave function in this basis may be chosen real:
\begin{equation}
    \psi_r(k)
    =
    \left(\frac{2V_r}{\pi}\right)^{1/4}
    e^{-V_r k^2}.
    \label{eq:arb-wavefunction}
\end{equation}
Indeed,
\begin{equation}
    \int_{\mathbb R}
    k\,|\psi_r(k)|^2\,dk
    =0,
\end{equation}
and
\begin{equation}
    \int_{\mathbb R}
    k^2|\psi_r(k)|^2\,dk
    =
    \frac{1}{4V_r},
\end{equation}

The Gaussian displacement noise acts diagonally in the displacement
variable and therefore produces the kernel
\begin{equation}
    \rho_0(k,k')
    =
    \psi_r(k)\psi_r(k')
    e^{-S(k-k')^2/2}.
    \label{eq:arb-rho-kernel}
\end{equation}
under $H_0$.
For $s\in\{+1,-1\}$, define the trace-class Gaussian operators
\begin{equation}
    X_s(k,k')
    :=
    \psi_r(k)\psi_r(k')
    \exp\!\left[
        -\frac{S}{2}(k-k'-s\omega)^2
    \right].
    \label{eq:arb-Xs-kernel}
\end{equation}

The kernel under $H_1$ gives
\begin{equation}
\begin{aligned}
    \rho_1(k,k')
    =
    \psi_r(k)\psi_r(k')
    \Big[
        &e^{-S(k-k')^2/2}
        -2i\epsilon_0
        e^{-S(k-k'-\omega)^2/2}
\\
        &\qquad
        +2i\epsilon_0
        e^{-S(k-k'+\omega)^2/2}
    \Big].
\end{aligned}
\label{eq:arb-rho1-kernel}
\end{equation}
Hence
\begin{equation}
    \Delta
    :=
    \rho_1-\rho_0
    =
    -2i\epsilon_0(X_+-X_-).
    \label{eq:arb-Delta}
\end{equation}

Introducing
\begin{equation}
    c:=e^{-S\omega^2/2},
    \qquad
    \kappa:=S\omega,
    \label{eq:arb-c-kappa}
\end{equation}
we may complete the square in the kernel of $X_s$:
\begin{align}
    -\frac{S}{2}(k-k'-s\omega)^2
    &=
    -\frac{S}{2}(k-k')^2
    +sS\omega(k-k')
    -\frac{S\omega^2}{2}
    \notag\\
    &=
    -\frac{S}{2}(k-k')^2
    +s\kappa(k-k')
    +\log c.
\end{align}
Since
\begin{equation}
    \bra{k}
    e^{s\kappa\hat K}\rho_0e^{-s\kappa\hat K}
    \ket{k'}
    =
    e^{s\kappa(k-k')}\rho_0(k,k'),
\end{equation}
we obtain the operator identity
\begin{equation}
    X_s
    =
    c\,e^{s\kappa\hat K}
    \rho_0
    e^{-s\kappa\hat K}.
    \label{eq:arb-similarity}
\end{equation}
Because $\rho_1=\rho_0+\Delta$ and
$\operatorname{Tr}\Delta=0$, we have the exact identity
\begin{align}
    \widetilde D_2(\rho_1\Vert\rho_0)
    &:=
    \log
    \operatorname{Tr}\!\left[
        \left(
            \rho_0^{-1/4}
            \rho_1
            \rho_0^{-1/4}
        \right)^2
    \right]
    \notag\\
    &=
    \log\!\left[
        1+
        \operatorname{Tr}\!\left(
            \Delta\rho_0^{-1/2}
            \Delta\rho_0^{-1/2}
        \right)
    \right].
    \label{eq:arb-D2-quadratic}
\end{align}
Thus, it remains to evaluate the quadratic form $    \operatorname{Tr}\!\left[
        \Delta\rho_0^{-1/2}
        \Delta\rho_0^{-1/2}
    \right]$

Using Eq.~\eqref{eq:arb-Delta}, this becomes
\begin{align}
    \operatorname{Tr}\!\left[
        \Delta\rho_0^{-1/2}
        \Delta\rho_0^{-1/2}
    \right]
    &=
    (-2i\epsilon_0)^2
    \operatorname{Tr}\!\left[
        (X_+-X_-)
        \rho_0^{-1/2}
        (X_+-X_-)
        \rho_0^{-1/2}
    \right]
    \notag\\
    &=
    -4\epsilon_0^2
    \left(
        T_{++}-T_{+-}-T_{-+}+T_{--}
    \right),
    \label{eq:arb-delta-trace-expanded}
\end{align}
where
\begin{equation}
    T_{st}
    :=
    \operatorname{Tr}\!\left[
        X_s\rho_0^{-1/2}
        X_t\rho_0^{-1/2}
    \right],
    \qquad
    s,t\in\{+1,-1\}.
    \label{eq:arb-Tst}
\end{equation}

The traces $T_{st}$ can be evaluated by exploiting the Gaussian
structure of $\rho_0$. Since $\rho_0$ is a centered one-mode Gaussian
state, there exists a Gaussian unitary $U_G$ that maps $\rho_0$ to a
thermal state. We use this unitary change of mode to evaluate
$T_{st}$ in the thermal basis. Importantly,
$e^{s\kappa\hat K}$ is not itself a Gaussian unitary, but under the
same change of mode it is represented as the exponential of a
quadrature of the thermal mode. The corresponding thermal-state
mapping is given in Appendix~\ref{app:thermal-mapping}. In particular, defining
\begin{equation}
    a:=V_r,
    \qquad
    \nu:=\frac{1}{2}\sqrt{\frac{V}{a}},
    \label{eq:arb-a-nu}
\end{equation}
the corresponding thermal state can be written as
\begin{equation}
    \rho_0
    =
    (1-q)q^{\hat n},
    \qquad
    q
    =
    \frac{\sqrt{V/a}-1}
         {\sqrt{V/a}+1}.
    \label{eq:arb-q}
\end{equation}
In the corresponding thermal mode $\hat b$, one has
\begin{equation}
    e^{\kappa\hat K}
    \longmapsto
    e^{z(\hat b+\hat b^\dagger)},
    \qquad
    z^2
    =
    \frac{\kappa^2\nu}{2V}.
    \label{eq:arb-z-definition}
\end{equation}
The explicit evaluation of $T_{st}$ is given in
Appendix~\ref{app:BCH-trace}, yielding
\begin{equation}
    T_{st}=c^2e^{-stM},
    \label{eq:arb-M-first}
\end{equation}
where 
\begin{equation}
    M
    =
    2z^2
    \left(q^{-1/2}-q^{1/2}\right) =
    \frac{S^{3/2}\omega^2}{\sqrt{V_r+S}}
    =
    S\omega^2\sqrt{\frac{S}{V}}.
\end{equation}

where $V=V_r+S$.

Substituting Eq.~\eqref{eq:arb-M-first} into Eq.~\eqref{eq:arb-delta-trace-expanded} and evaluating the resulting quadratic form as given in Appendix~\ref{app:quadratic} gives
\begin{align}
    \operatorname{Tr}\!\left[
        \Delta\rho_0^{-1/2}
        \Delta\rho_0^{-1/2}
    \right]
    &=
    16\epsilon_0^2\mathcal R,
    \label{eq:arb-delta-trace-expanded2}
\end{align}

where
\begin{equation}
    \mathcal R
    :=
    e^{-S\omega^2}
    \sinh\!\left(
        S\omega^2\sqrt{\frac{S}{V}}
    \right).
    \label{eq:arb-R}
\end{equation}
Therefore 
\begin{align}
    \widetilde D_2(\rho_1\Vert\rho_0)  
    &=
    \log\!\left(
        1+16\epsilon_0^2\mathcal R
    \right).
    \label{eq:arb-D2}
\end{align}

We now consider an arbitrary receiver measurement. Let $P_0$ and
$P_1$ denote the outcome distributions generated by an arbitrary
POVM applied to the $N$ output states. Such a measurement, including
a collective POVM and any hypothesis-independent ancilla, is a
quantum channel from the output system to a classical register.
Therefore, by the data-processing inequality for the sandwiched
Rényi divergence,
\begin{equation}
    D_2(P_1\Vert P_0)
    \leq
    \widetilde D_2
    \left(
        \rho_1^{\otimes N}
        \middle\Vert
        \rho_0^{\otimes N}
    \right).
    \label{eq:arb-DPI-general}
\end{equation}

Since the sandwiched order-two Rényi divergence is additive on tensor
products,
\begin{equation}
    \widetilde D_2
    \left(
        \rho_1^{\otimes N}
        \middle\Vert
        \rho_0^{\otimes N}
    \right)
    =
    N\widetilde D_2(\rho_1\Vert\rho_0).
    \label{eq:arb-D2-additivity}
\end{equation}
Consequently,
\begin{equation}
    \log\!\left(
        1+\chi^2(P_1\Vert P_0)
    \right)
    \leq
    N
    \log\!\left(
        1+16\epsilon_0^2\mathcal R
    \right).
    \label{eq:arb-DPI}
\end{equation}

For equal priors, a success probability of at least $2/3$ requires
$\operatorname{TVD}(P_0,P_1)\geq1/3$. Using
$\operatorname{TVD}(P_0,P_1)\leq
\frac{1}{2}\sqrt{\chi^2(P_1\Vert P_0)}$ therefore gives
$\chi^2(P_1\Vert P_0)\geq4/9$, or equivalently
\begin{equation}
    D_2(P_1\Vert P_0)
    =
    \log\!\left[1+\chi^2(P_1\Vert P_0)\right]
    \geq
    \log\!\left(\frac{13}{9}\right).
    \label{eq:arb-classical-D2-lower}
\end{equation}

Combining Eqs.~\eqref{eq:arb-DPI-general},
\eqref{eq:arb-D2-additivity}, and
\eqref{eq:arb-classical-D2-lower} gives
\begin{equation}
    N\widetilde D_2(\rho_1\Vert\rho_0)
    \geq
    \log\!\left(\frac{13}{9}\right).
    \label{eq:arb-N-D2}
\end{equation}
Using Eq.~\eqref{eq:arb-D2}, we obtain the exact bound
\begin{equation}
    N
    \geq
    \frac{\log(13/9)}
    {\log\!\left(1+16\epsilon_0^2\mathcal R\right)}.
    \label{eq:arb-exact-lower-bound}
\end{equation}

To expose the exponential dependence, Eq.~\eqref{eq:arb-R} gives
$\mathcal R\leq
\frac{1}{2}e^{-S\omega^2(1-\sqrt{S/V})}$.
Using $\log(1+x)\leq x$ in
Eq.~\eqref{eq:arb-exact-lower-bound}, we therefore obtain
\begin{equation}
    N
    \geq
    \frac{\log(13/9)}
    {8\epsilon_0^2}
    \exp\!\left[
        S\omega^2
        \left(1-\sqrt{\frac{S}{V}}\right)
    \right].
    \label{eq:arb-intermediate-bound}
\end{equation}

Finally, substituting
$S=1/(4\sigma_\beta^2)$,
$\omega=2|\beta_0|$, and
$V=V_r+1/(4\sigma_\beta^2)$ gives
$S\omega^2=|\beta_0|^2/\sigma_\beta^2$ and
$\sqrt{S/V}=
[1+4V_r\sigma_\beta^2]^{-1/2}$. Hence,
\begin{equation}
    N\geq
    \frac{\log(13/9)}
    {8\epsilon_0^2}
    \exp\!\left[
        \frac{|\beta_0|^2}{\sigma_\beta^2}
        \left(
            1-
            \frac{1}
            {\sqrt{1+4V_r\sigma_\beta^2}}
        \right)
    \right].
    \label{eq:arb-simple-lower-bound}
\end{equation}

This proves Theorem~1.
\hfill$\square$

\begin{corollary}
The lower bounds of Theorem~1 with $V_r=1$ apply to every
optical-classical probe state, including arbitrary classically correlated
multimode probes and arbitrary collective output measurements.

More precisely, let the $N$-mode input state admit a nonnegative
Glauber--Sudarshan representation \cite{sudarshan1963equivalence}
\begin{equation}
 \rho_{\mathrm{cl}}^{(N)}
 =
 \int_{\mathbb C^N}
 P_N(\boldsymbol{\alpha})\,
 D(\boldsymbol{\alpha})
 \ket{0}\!\bra{0}^{\otimes N}
 D^\dagger(\boldsymbol{\alpha})\,
 d^{2N}\boldsymbol{\alpha},
 \qquad
 P_N(\boldsymbol{\alpha})\geq0.
 \label{eq:classical-P-state}
\end{equation}
Then any protocol using $\rho_{\mathrm{cl}}^{(N)}$ and succeeding with
probability at least $2/3$ requires
\begin{equation}
 N\geq
 \frac{\log(13/9)}{8\epsilon_0^2}
 \exp\!\left[
 \frac{|\beta_0|^2}{\sigma_\beta^2}
 \left(
 1-\frac{1}{\sqrt{1+4\sigma_\beta^2}}
 \right)
 \right].
 \label{eq:classical-simple-lower-bound}
\end{equation}
\paragraph{Proof.}
Random-displacement channels are displacement covariant:
\begin{equation}
 \Lambda_j\!\left(
 D(\alpha)\rho D^\dagger(\alpha)
 \right)
 =
 D(\alpha)\Lambda_j(\rho)D^\dagger(\alpha).
 \label{eq:displacement-covariance}
\end{equation}
Let $\rho_j^{\mathrm{vac}}$ denote the single-use output state in
Eq.~\eqref{eq:arb-output-states} with $V_r=1$.  It follows from
Eqs.~\eqref{eq:classical-P-state} and
\eqref{eq:displacement-covariance} that
\begin{equation}
 \Lambda_j^{\otimes N}\!\left(\rho_{\mathrm{cl}}^{(N)}\right)
 =
 \mathcal C_{P_N}
 \left[
 \left(\rho_j^{\mathrm{vac}}\right)^{\otimes N}
 \right],
 \label{eq:classical-output-reduction}
\end{equation}
where
\begin{equation}
 \mathcal C_{P_N}(X)
 :=
 \int_{\mathbb C^N}
 P_N(\boldsymbol{\alpha})\,
 D(\boldsymbol{\alpha})X
 D^\dagger(\boldsymbol{\alpha})\,
 d^{2N}\boldsymbol{\alpha}
\end{equation}
is a completely positive trace-preserving map independent of the
hypothesis $j$.

Data processing of the sandwiched order-two R\'enyi divergence therefore
gives
\begin{align}
 \widetilde D_2\!\left(
 \Lambda_1^{\otimes N}(\rho_{\mathrm{cl}}^{(N)})
 \middle\Vert
 \Lambda_0^{\otimes N}(\rho_{\mathrm{cl}}^{(N)})
 \right)
 &\leq
 \widetilde D_2\!\left(
 (\rho_1^{\mathrm{vac}})^{\otimes N}
 \middle\Vert
 (\rho_0^{\mathrm{vac}})^{\otimes N}
 \right)\\
 &=N\log\!\left(
 1+16\epsilon_0^2\mathcal R_{\mathrm{vac}}
 \right).
\end{align}
The remainder of the argument is identical to
Eqs.~\eqref{eq:arb-DPI}--\eqref{eq:arb-simple-lower-bound}.
\hfill$\square$\\
\end{corollary}

\section{Minimum Achievable Sample Complexity from Homodyne Measurements}
\label{sec:LLR-lower-bound}

\subsection{LLR Decision based on output distribution}
Theorem~1 establishes a lower bound
on the sample complexity for a fixed squeezed-vacuum probe even when the
receiver is allowed to perform an arbitrary collective POVM on all output
states. We now consider the experimentally relevant case in which the
receiver performs homodyne measurements of a fixed quadrature on each
channel output. The resulting inference problem is purely classical: each channel use produces one independent real-valued sample drawn from one of two output probability distributions.

This restriction allows us to determine the sample complexity more
explicitly through the log-likelihood ratio (LLR). In particular, we
derive the mean and variance of the single-sample LLR and use the central
limit theorem to obtain the number of homodyne samples required to achieve
a prescribed discrimination success probability. Unlike Theorem~1, which
allows arbitrary collective measurements, the result in this section
describes the performance of the optimal decision rule for the specified
homodyne measurement.

\subsection*{Homodyne output distributions}

For a fixed quadrature measurement, the output distributions are obtained
by convolving the displacement distribution with the probe quadrature
distribution. Under $H_0$,
\begin{equation}
    p_{0,\mathrm{out}}(x)
    =
    \frac{1}{\sqrt{2\pi V}}
    e^{-x^2/(2V)},
    \qquad
    V=\frac{1}{4\sigma_\beta^2}+\sigma_Q^2,
    \label{eq:LLR-p0out}
\end{equation}
where $\sigma_Q^2$ is the variance of the measured quadrature. Under
$H_1$, the Gaussian convolution attenuates and broadens the oscillatory
component, giving
\begin{equation}
    p_{1,\mathrm{out}}(x)
    =
    p_{0,\mathrm{out}}(x)
    \left[1+4\epsilon_0A\sin(kx)\right],
    \label{eq:LLR-p1out}
\end{equation}
where
\[
    k=\frac{2\beta_0}{1+4\sigma_\beta^2\sigma_Q^2},
    \qquad
    A=
    \exp\!\left[
        -\frac{2\beta_0^2\sigma_Q^2}
        {1+4\sigma_\beta^2\sigma_Q^2}
    \right].
\]
Unlike Theorem~1, which allows arbitrary measurements of the full
quantum output state, this homodyne analysis fixes the measured
quadrature and therefore depends only on its variance $\sigma_Q^2$.

\subsection*{Log-likelihood Ratio}

For $N$ independent homodyne samples, the Neyman--Pearson optimal
decision rule is based on
\begin{equation}
    \mathrm{LLR}
    =
    \sum_{j=1}^N
    \log\!\left[1+4\epsilon_0A\sin(kx_j)\right].
    \label{eq:LLR-total}
\end{equation}
For equal priors, $H_1$ is selected when $\mathrm{LLR}>0$ and $H_0$
otherwise, giving
\begin{equation}
    P_{\mathrm{succ}}
    =
    \frac12 P(\mathrm{LLR}>0\mid H_1)
    +
    \frac12 P(\mathrm{LLR}\leq0\mid H_0).
    \label{eq:LLR-success}
\end{equation}
The moments of the LLR are then expanded perturbatively in
$\epsilon_0$; the resulting expressions determine its asymptotic
distribution and sample complexity.

\subsection*{Approximate Mean of the LLR}

Using
$\log(1+u)=u-u^2/2+O(u^3)$ with
$u=4\epsilon_0A\sin(kx)$, the single-sample mean under $H_0$ is
\begin{align}
    \mu_0
    &:=
    \mathbb{E}_{H_0}[\ell(x)]
    \notag\\
    &=
    4\epsilon_0A\,\mathbb{E}_{H_0}[\sin(kx)]
    -8\epsilon_0^2A^2\,
    \mathbb{E}_{H_0}[\sin^2(kx)]
    +O(\epsilon_0^3).
\end{align}
Because $p_{0,\mathrm{out}}(x)$ is centered and symmetric,
$\mathbb{E}_{H_0}[\sin(kx)]=0$. Moreover,
$\sin^2(kx)=[1-\cos(2kx)]/2$, and for a centered Gaussian with
variance $V$,
$\mathbb{E}_{H_0}[\cos(2kx)]=e^{-2k^2V}$. Hence
\begin{equation}
    \mu_0
    =
    -4\epsilon_0^2A^2
    \left(1-e^{-2k^2V}\right)
    +O(\epsilon_0^3).
    \label{eq:LLR-mu0}
\end{equation}
The negative sign reflects the vanishing first-order contribution and
the concavity of $\log(1+u)$.

\subsection*{Approximate Mean of the LLR under $H_1$}

Under $H_1$,
$p_{1,\mathrm{out}}(x)
=p_{0,\mathrm{out}}(x)[1+4\epsilon_0A\sin(kx)]$.
Therefore,
\begin{align}
    \mu_1
    &:=
    \mathbb{E}_{H_1}[\ell(x)]
    \notag\\
    &=
    \mu_0+
    4\epsilon_0A\,
    \mathbb{E}_{H_0}[\ell(x)\sin(kx)].
\end{align}
To leading order,
$\ell(x)=4\epsilon_0A\sin(kx)+O(\epsilon_0^2)$, so using
$\mathbb{E}_{H_0}[\sin^2(kx)]
=[1-e^{-2k^2V}]/2$ gives
\begin{equation}
    \mu_1
    =
    4\epsilon_0^2A^2
    \left(1-e^{-2k^2V}\right)
    +O(\epsilon_0^3).
    \label{eq:LLR-mu1}
\end{equation}
Thus, to the order retained, $\mu_1=-\mu_0$. This is an
$O(\epsilon_0^2)$ relation rather than an exact identity.

\subsection*{Approximate Variance of the Single-Sample LLR}

To leading order,
$\ell(x)=4\epsilon_0A\sin(kx)+O(\epsilon_0^2)$.
Since $\mathbb{E}_{H_0}[\sin(kx)]=0$,
\begin{equation}
    \sigma_{\ell,0}^2
    :=
    \operatorname{Var}_{H_0}[\ell(x)]
    =
    8\epsilon_0^2A^2
    \left(1-e^{-2k^2V}\right)
    +O(\epsilon_0^3).
    \label{eq:LLR-var0}
\end{equation}
The same leading-order expression holds under $H_1$,
\begin{equation}
    \sigma_{\ell,1}^2
    =
    8\epsilon_0^2A^2
    \left(1-e^{-2k^2V}\right)
    +O(\epsilon_0^3).
    \label{eq:LLR-var1}
\end{equation}

It is therefore convenient to define
\begin{equation}
    \Omega
    :=
    4\epsilon_0^2A^2
    \left(1-e^{-2k^2V}\right),
    \label{eq:LLR-Omega}
\end{equation}
so that, to the order considered,
\begin{equation}
    \mu_1=\Omega,\qquad
    \mu_0=-\Omega,\qquad
    \sigma_{\ell,0}^2
    =
    \sigma_{\ell,1}^2
    =
    2\Omega.
    \label{eq:LLR-moments}
\end{equation}

\subsection*{Asymptotic Distribution of the Total LLR}

The total LLR is a sum of $N$ independent single-sample contributions.
For sufficiently large $N$, the central limit theorem gives
\begin{align}
    \mathrm{LLR}\mid H_1
    &\overset{\mathrm{CLT}}{\sim}
    \mathcal{N}(N\Omega,2N\Omega),\\
    \mathrm{LLR}\mid H_0
    &\overset{\mathrm{CLT}}{\sim}
    \mathcal{N}(-N\Omega,2N\Omega).
\end{align}
Thus, with $\sigma_L=\sqrt{2N\Omega}$,
$P(\mathrm{LLR}>0\mid H_1)
=P(\mathrm{LLR}\leq0\mid H_0)
=\Phi\!\left(\sqrt{N\Omega/2}\right)$, and hence
\begin{equation}
    P_{\mathrm{succ}}
    =
    \Phi\!\left(\sqrt{\frac{N\Omega}{2}}\right).
    \label{eq:LLR-success-final}
\end{equation}

\subsection*{Asymptotic Sample Complexity for a Target Success Probability}

For equal priors, requiring $P_{\mathrm{succ}}\geq2/3$ gives
$\sqrt{N\Omega/2}\geq z_*$, where
$z_*:=\Phi^{-1}(2/3)\approx0.4307$. Therefore,
\begin{equation}
    N\geq\frac{2z_*^2}{\Omega}.
    \label{eq:LLR-N-Omega}
\end{equation}
Using Eq.~\eqref{eq:LLR-Omega}, this becomes
\begin{equation}
    N
    \geq
    \frac{z_*^2}
    {2\epsilon_0^2A^2
    \left(1-e^{-2k^2V}\right)}.
    \label{eq:LLR-N-A}
\end{equation}
Since
$A^2=\exp[-4\beta_0^2\sigma_Q^2/
(1+4\sigma_\beta^2\sigma_Q^2)]$, the approximate minimum sample
complexity is
\begin{equation}
    N_{\min}
    \approx
    \frac{z_*^2}
    {2\epsilon_0^2
    \left(1-e^{-2k^2V}\right)}
    \exp\!\left[
        \frac{4\beta_0^2\sigma_Q^2}
        {1+4\sigma_\beta^2\sigma_Q^2}
    \right].
    \label{eq:LLR-N-final}
\end{equation}
Numerically, $z_*^2\simeq0.1855$, giving
$N_{\min}\approx
0.0928[\epsilon_0^2(1-e^{-2k^2V})]^{-1}
\exp[4\beta_0^2\sigma_Q^2/
(1+4\sigma_\beta^2\sigma_Q^2)]$.

\subsection*{Limiting behavior}

\paragraph{Large-$\beta_0$ limit.}

As $\beta_0$ increases, $k^2V\rightarrow\infty$, so
$1-e^{-2k^2V}\rightarrow1$. Equation~\eqref{eq:LLR-N-final} therefore
reduces to
\begin{equation}
    N_{\min}
    \approx
    \frac{z_*^2}{2\epsilon_0^2}
    \exp\!\left[
        \frac{4\beta_0^2\sigma_Q^2}
        {1+4\sigma_\beta^2\sigma_Q^2}
    \right].
    \label{eq:LLR-large-beta}
\end{equation}
Thus, the sample complexity is exponentially sensitive to $\beta_0^2$,
with the exponent controlled by the measured quadrature variance
$\sigma_Q^2$. Reducing this variance suppresses the exponential penalty
associated with resolving high-frequency features.

\paragraph{Small-$\beta_0$ limit.}

As $\beta_0\rightarrow0$, $A\rightarrow1$ and $k\rightarrow0$, so
$1-e^{-2k^2V}\sim2k^2V\rightarrow0$ and consequently
$\Omega\rightarrow0$. Hence $N_{\min}\rightarrow\infty$, consistent
with the two output distributions becoming identical as
$\beta_0\rightarrow0$.
\subsection{Sample complexity using characteristic-function reconstruction}
\label{sec:homodyne-achievable}

We now derive an achievable sample complexity for reconstructing the
characteristic function from homodyne measurements. For a Gaussian probe with measured quadrature variance $\sigma_Q^2$, a
homodyne outcome is $Y_k=X_k+Q_k$, where
$Q_k\sim\mathcal N(0,\sigma_Q^2)$. The corresponding noise
characteristic function is
$\lambda_Q(\beta)=\mathbb E[e^{2i\beta Q}]
=e^{-2\sigma_Q^2\beta^2}$, so that
$\lambda_Y(\beta)=\lambda(\beta)\lambda_Q(\beta)$. Hence, from $N$
independent outcomes, the unbiased deconvolution estimator is
\begin{equation}
    \tilde{\lambda}(\beta)
    =
    e^{2\sigma_Q^2\beta^2}
    \frac{1}{N}\sum_{k=1}^N e^{2i\beta Y_k}.
    \label{eq:hom-deconv-estimator}
\end{equation}

We consider reconstruction at the bandwidth edge $\beta=\beta_0$, which
is the most demanding frequency. The distinguishing threshold is
\begin{equation}
    \epsilon
    =
    \frac{1}{2}\operatorname{Im}[\lambda_1(\beta_0)]
    =
    \epsilon_0
    \left(
        1-e^{-2\beta_0^2/\sigma_\beta^2}
    \right).
    \label{eq:hom-epsilon}
\end{equation}

\subsubsection*{Imaginary-part estimator}

Writing
$e^{2i\beta_0Y_k}=\cos(2\beta_0Y_k)+i\sin(2\beta_0Y_k)$, define
$I_k=\sin(2\beta_0Y_k)\in[-1,1]$. Then
\[
\operatorname{Im}\tilde{\lambda}(\beta_0)
=
e^{2\sigma_Q^2\beta_0^2}\frac{1}{N}\sum_{k=1}^N I_k,
\]
and, with $\mu_I=\mathbb E[I_k]$,
\begin{equation}
    \operatorname{Im}\tilde{\lambda}(\beta_0)
    -
    \operatorname{Im}\lambda(\beta_0)
    =
    e^{2\sigma_Q^2\beta_0^2}
    \left(
        \frac{1}{N}\sum_{k=1}^N I_k-\mu_I
    \right).
    \label{eq:hom-imag-error}
\end{equation}

Requiring the reconstruction error to exceed $\epsilon$ with probability
at most $\delta$ is therefore equivalent to bounding the sample-mean
deviation at the rescaled threshold
$\epsilon e^{-2\sigma_Q^2\beta_0^2}$.

\subsubsection*{Hoeffding bound}

Since $I_k\in[-1,1]$ are independent, Hoeffding's inequality gives \cite{hoeffding1963probability}
\[
\Pr\!\left(
\left|\frac{1}{N}\sum_{k=1}^N I_k-\mu_I\right|\geq t
\right)
\leq
2e^{-Nt^2/2}.
\]
Setting $t=\epsilon e^{-2\sigma_Q^2\beta_0^2}$ yields
\begin{equation}
    N
    \geq
    \frac{2\log(2/\delta)}
         {\epsilon^2}
    e^{4\sigma_Q^2\beta_0^2}.
    \label{eq:hom-hoeffding-bound}
\end{equation}
which is a sufficient condition for the estimator error probability to be at most $\delta$. For $\delta=1/3$, this becomes
$N\geq2\log(6)\epsilon^{-2}e^{4\sigma_Q^2\beta_0^2}$.
Substituting Eq.~\eqref{eq:hom-epsilon}, we obtain
\begin{equation}
    N
    \geq
    \frac{2\log(2/\delta)}
    {\epsilon_0^2
    \left(1-e^{-2\beta_0^2/\sigma_\beta^2}\right)^2}
    \exp\!\left(4\sigma_Q^2\beta_0^2\right).
    \label{eq:hom-hoeffding-final}
\end{equation}

This establishes an achievable upper bound for estimating the
distinguishing component of the characteristic function. The exponential
factor $\exp(4\sigma_Q^2\beta_0^2)$ arises from the inversion of the
Gaussian measurement-noise characteristic function.

\subsubsection*{Vacuum versus squeezed probes}

For a vacuum probe,
\begin{equation}
    \sigma_{Q,\mathrm{vac}}^2=1.
\end{equation}
For a squeezed probe with transmission $T$ and squeezing parameter
$r$,
\begin{equation}
    \sigma_{Q,\mathrm{sq}}^2
    =
    (1-T)+Te^{-2r}.
    \label{eq:hom-effective-variance}
\end{equation}
Since the prefactors in Eq.~\eqref{eq:hom-hoeffding-final} are identical
apart from the dependence through $\epsilon$ and the finite-frequency
variance correction, the leading exponential improvement is
\begin{equation}
    \frac{N_{\mathrm{vac}}}{N_{\mathrm{sq}}}
    \sim
    \exp\!\left[
        4
        \left(
            1-\sigma_{Q,\mathrm{sq}}^2
        \right)
        \beta_0^2
    \right]
\end{equation}
in the large-$\beta_0$ regime.

Thus, for any effective squeezed-quadrature variance satisfying
$\sigma_{Q,\mathrm{sq}}^2<1$, homodyne detection provides an exponential
reduction in the number of samples required to resolve the high-frequency
feature.

It is important to distinguish this achievable scaling from the
measurement-independent lower bound of
Theorem~1. The latter allows arbitrary
measurements on the full quantum output state and therefore characterizes
a more general task. The present result is restricted to homodyne
detection and provides an explicit achievable scaling for the
experimentally implemented measurement strategy.

\section{Experimental Implementation Details}
\subsection{State Preparation}
\begin{figure}[h]
    \centering
    \includegraphics[width=0.8\columnwidth]{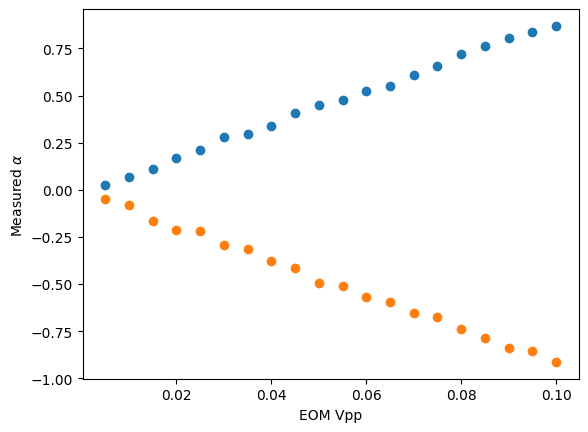}
    \caption{\textbf{EOM linearity characterization.} Measured displacement 
    amplitude $\alpha$ as a function of EOM peak-to-peak voltage $V_\text{pp}$ 
    for positive (blue) and negative (orange) displacement directions. The linear 
    response confirms that the EOM operates in the small-modulation regime 
    throughout the experimental range, validating the sideband amplitude 
    approximation used in the state preparation protocol.}
    \label{fig:eom_linearity}
\end{figure}

To produce displaced squeezed states experimentally, a squeezed vacuum state is combined with a displaced coherent beam on a highly reflective (HR) beamsplitter with reflectivity $\eta = 98:2$. The auxiliary coherent beam of power $P_\text{aux}$
 introduces a displacement while adding negligible additional noise, so that the variance of the output state remains that of the squeezed vacuum and the mean displacement becomes 
\begin{equation}\alpha_\text{eff} \approx \sqrt{1-T}\,\alpha_\text{aux}\end{equation} 
A phase modulator (EOM) driven at frequency $\Omega = 1$ MHz produces coherent sidebands in the small-modulation regime, where the sideband amplitude scales linearly with the applied voltage. The linearity of the EOM response is verified across the full experimental voltage range, confirming operation in the small-modulation regime throughout (Fig.~\ref{fig:eom_linearity}).
In the experiments reported here, we fix $\sigma_\beta = 1$, so that the real-space envelope width $1/(2\sigma_\beta) = 0.5$ is expressed in shot-noise units (SNU), consistent with how the displacement $\alpha$ is calibrated experimentally against the shot-noise variance, and vary the peak separation $\beta_0$ over the range [0.5, 6] relative to this scale. This is not an experimental parameter being tuned but a choice of normalization for how the hypothesis pair is defined, and the same underlying experiment is described regardless of this choice. Fixing $\sigma_\beta$ at the shot-noise scale keeps the two hypotheses equally energetic, ensuring the distinguishing difficulty arises purely from the fringe structure rather than from any energy imbalance between $H_0$ and $H_1$.
\subsection{Error Characterization}

\begin{figure}[h]
    \centering
    \includegraphics[width=\columnwidth]{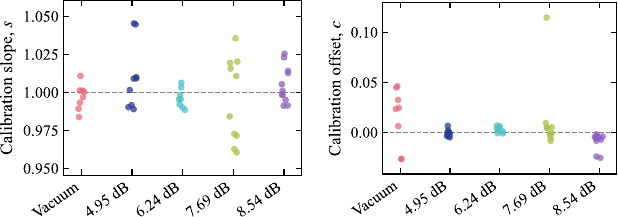}
    \caption{
    \textbf{Displacement calibration parameters.}
    Affine calibration parameters extracted independently for each
    experimental data set. Each point corresponds to one value of
    $\beta_0$; points are grouped according to the squeezing level.
    (a) Fitted calibration slope $s$, with the dashed line indicating the
    ideal value $s=1$. (b) Fitted calibration offset $c$, with the dashed
    line indicating $c=0$. The observed deviations are corrected on a
    data-set-by-data-set basis using $Y'=(Y-c)/s$ before
    characteristic-function estimation.
    }
    \label{fig:alpha_calibration}
\end{figure}

In practice the applied displacement and the calibrated noise model are both subject to experimental imperfections. We therefore distinguish between errors in the displacement calibration and uncertainty in the noise variance used for deconvolution. The homodyne outcome is more accurately modeled as
\begin{equation}Y = X + \delta x + Q + \delta Q\end{equation}
where $\delta x$ represents a systematic miscalibration between the applied displacement and the displacement actually realized on the state (e.g. modulator nonlinearity, detector gain, path loss), and $\delta Q$
represents an error in the assumed noise model itself, i.e. an uncertainty in $\sigma_Q^2$ relative to its true value. In our setup, $\delta x$ is well described by an affine distortion, $X_{\rm meas} = s\,X + c$, with s,c obtained from a linear fit of measured against applied displacement across the full data set. Since this component of the error is deterministic and 
X-dependent rather than stochastic, it is exactly invertible: defining $Y' \equiv (Y-c)/s$ removes the bias, at the cost of rescaling the residual noise variance entering the deconvolution kernel by $\sigma_Q^2/s^2$. The fitted calibration parameters remain close to their ideal values over the experimental data sets, with $s\in[0.98,1.04]$ and offsets much smaller than one SNU. Figure~\ref{fig:alpha_calibration} shows a representative calibration over the full displacement range. All experimental data are corrected using the corresponding fitted values of $s$ and $c$ prior to characteristic-function estimation.
The remaining term, $\delta Q$, is more delicate. Writing the assumed noise variance used in the deconvolution as $\sigma_{Q,\rm used}^2 = \sigma_Q^2 + \delta Q$, the reconstructed characteristic function acquires a multiplicative error
\begin{equation}\tilde\lambda(\beta) \approx \lambda(\beta)\, e^{2\,\delta Q\, \beta^2}\end{equation}
which follows directly from the deconvolution relation 
\begin{equation}\tilde\lambda(\beta) = \lambda_Y(\beta)/\lambda_Q(\beta)\end{equation}
once \begin{equation}\lambda_Q(\beta) = e^{-2\sigma_{Q,\rm used}^2\beta^2}\end{equation} is evaluated at the wrong variance. Because this error enters as $e^{2\delta Q \beta^2}$, it grows exponentially with $\beta^2$ precisely the same exponential dependence that governs the sample-complexity. So even a small mismatch $\delta Q$ becomes the dominant source of error at the bandwidth edge $\beta \sim \beta_0$ relevant to distinguishing $H_0$ from $H_1$. Accurate calibration of $\sigma_Q^2$ is therefore at least as important as the sample size itself in realizing the exponential advantage predicted for the squeezed probe.

\section{Experimental data processing}
\subsection{Independent Sample Construction}
A key requirement for estimating the success probability as a function of
the sample number $N$ is to reduce temporal correlations in the homodyne
records. Consecutive measurements are not perfectly independent because
of finite temporal correlations in the detection chain. We therefore
characterize the temporal correlations using the measured normalized
autocorrelation function. The correlation time, defined by the $1/e$
crossing, is approximately
\begin{equation}
    \tau_c \simeq 0.72~\mu\mathrm{s}.
\end{equation}
At the sampling rate of $4.17$~MHz, this corresponds to approximately
three samples. We therefore group consecutive measurements into
non-overlapping blocks of three samples and average within each block.
This produces an effective set of samples with substantially reduced
temporal correlations. After this procedure, each homodyne trace
contains approximately
\begin{equation}
    N_{\mathrm{eff}}
    =
    \frac{T}{\tau_c}
    \simeq 1.67\times10^3
\end{equation}
effective samples.

For each displacement amplitude, the resulting grouped samples form a
pool from which reconstruction datasets are generated. A dataset is
constructed by randomly selecting one grouped sample for each
displacement amplitude. Distinct index combinations are used when
generating the ensemble of datasets, so that the same complete dataset
realization is not repeated. We generate 2000 such dataset realizations.
The datasets are normalized to shot-noise units and corrected for the
measured affine displacement distortion prior to reconstruction.

A null dataset collected at $\beta_0=0$ under otherwise identical
experimental conditions serves as the reference for $H_0$ at each
squeezing level. For each $\beta_0>0$, the corresponding alternative
datasets represent $H_1$. The two hypotheses are selected with equal
prior probability.

For a given sample number $N$, one dataset is randomly selected from the
corresponding experimental ensemble, and $N$ effective samples are
randomly drawn from that dataset without replacement. For the hypothesis-testing and reconstruction tasks considered below,the success probability is estimated from $R=2000$ random trials. In each
trial, an experimental dataset is randomly selected from the corresponding
ensemble, and $N$ effective samples are randomly drawn from that dataset
without replacement. The precise success criterion depends on the task:
for binary hypothesis testing, $I_r(N)=1$ when the hypothesis is correctly
classified, while for reconstruction, $I_r(N)=1$ when the prescribed
reconstruction-error criterion is satisfied. The empirical success
probability is therefore
\begin{equation}
    \hat P_{\mathrm{succ}}(N)
    =
    \frac{1}{R}
    \sum_{r=1}^{R} I_r(N),
\end{equation}
where $I_r(N)\in\{0,1\}$ denotes success in the $r$th trial.
Since each outcome is binary, the statistical uncertainty in
$\hat P_{\mathrm{succ}}(N)$ is described by a binomial distribution,
with standard error
\begin{equation}
    \sigma_P(N)
    =
    \sqrt{
    \frac{\hat P_{\mathrm{succ}}(N)
    [1-\hat P_{\mathrm{succ}}(N)]}{R}
    }.
\end{equation}
For $R=2000$, this gives
$\sigma_P\simeq0.0105$ at the target probability
$P_{\mathrm{succ}}=2/3$.

The minimum sample complexity for hypothesis testing is defined as the
smallest sample number for which the estimated success probability
reaches the target value,
\begin{equation}
    N_{\min}
    =
    \min\left\{
    N:\hat P_{\mathrm{succ}}(N)\geq\frac{2}{3}
    \right\}.
\end{equation}

\subsection{Full-bandwidth characteristic-function reconstruction}

We experimentally validate the reconstruction protocol at the endpoint
$\beta_0$ and subsequently extend it across the accessible bandwidth.
The characteristic function is estimated from $N$ effective homodyne
samples using
\begin{equation}
    \widetilde{\lambda}(\beta)
    =
    e^{2V_{\mathrm{eff}}\beta^2}
    \frac{1}{N}
    \sum_{k=1}^{N}e^{2i\beta x_k},
    \label{eq:estimator}
\end{equation}
where $V_{\mathrm{eff}}$ is the experimentally determined effective
quadrature variance.

\begin{figure*}[t]
    \centering
    \includegraphics[width=\textwidth]{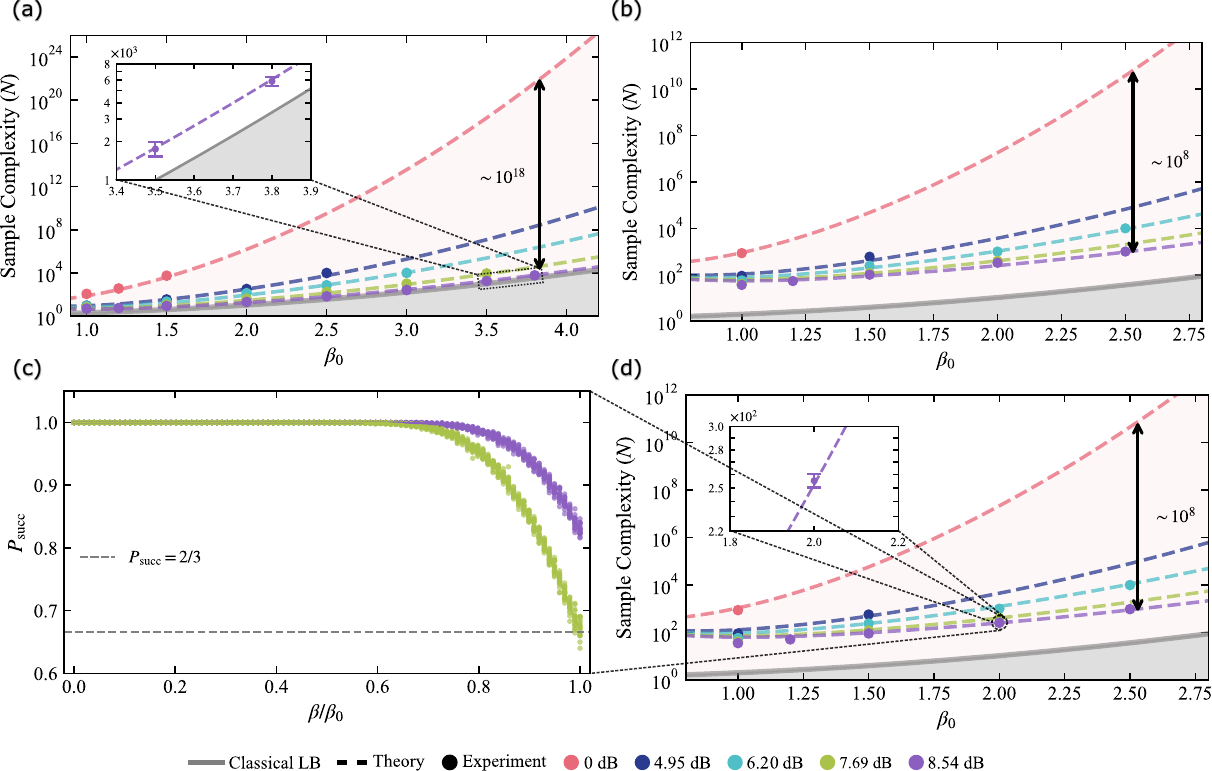}
    \caption{
    \textbf{Experimental validation of characteristic-function reconstruction
    and hypothesis testing.}
    (a) Sample complexity for binary hypothesis testing using the reconstructed
    characteristic-function value at the endpoint $\beta_0$.
    (b) Sample complexity for reconstruction of the characteristic function
    at $\beta_0$, with a reconstruction trial counted as successful when
    $|\widetilde{\lambda}(\beta_0)-\lambda(\beta_0)|\leq\epsilon$ and the
    target success probability $P_{\mathrm{succ}}=2/3$.
    (c) Pointwise reconstruction success probability as a function of
    $\beta/\beta_0$ at a fixed sample number $N$ chosen according to the
    Hoeffding bound for the full-bandwidth reconstruction protocol. The
    success probability is evaluated at 100 values of $\beta$ spanning
    $[0,\beta_0]$.
    (d) Sample complexity for simultaneous reconstruction at 25 sampled
    frequencies spanning $[0,\beta_0]$, with 24 frequencies drawn uniformly
    and $\beta_0$ included explicitly. A reconstruction trial is successful
    only when
    $\max_j|\widetilde{\lambda}(\beta_j)-\lambda(\beta_j)|\leq\epsilon$.
    The grey curves show the classical lower bound, while the dashed curves
    show the theoretical reconstruction bound and the markers show the
    experimental results.
    }
    \label{fig:SM_reconstruction}
\end{figure*}

We first consider reconstruction at the endpoint $\beta_0$, which is
the most demanding point for the estimator because of the exponential
prefactor in Eq.~\eqref{eq:estimator}. For a given $N$, $N$ effective
homodyne samples are randomly selected from an experimental dataset and
used to estimate $\widetilde{\lambda}(\beta_0)$. The reconstruction error
is defined as
\begin{equation}
    \Delta\lambda(\beta_0)
    =
    \left|
    \widetilde{\lambda}(\beta_0)
    -
    \lambda(\beta_0)
    \right|.
\end{equation}
A reconstruction trial is considered successful when
\begin{equation}
    \Delta\lambda(\beta_0)\leq\epsilon .
\end{equation}
The minimum sample number is determined as the smallest $N$ for which
the reconstruction success probability reaches
$P_{\mathrm{succ}}=2/3$. The resulting endpoint reconstruction sample
complexity is shown in Fig.~\ref{fig:SM_reconstruction}(b).

The reconstructed characteristic-function value at $\beta_0$ can
further be used directly for the binary hypothesis-testing task. In each
trial, the null and alternative hypotheses are selected with equal prior
probability, and $N$ effective homodyne samples are used to reconstruct
$\widetilde{\lambda}(\beta_0)$. The hypothesis is then assigned according
to the value of the imaginary part of the reconstructed characteristic
function relative to the decision threshold. The corresponding
experimental sample complexity is shown in
Fig.~\ref{fig:SM_reconstruction}(a).

To test reconstruction beyond a single frequency, we additionally
sample $24$ values of $\beta$ independently and uniformly from
$[0,\beta_0]$ and append $\beta_0$ explicitly, giving $25$ sampled
frequencies in total. For each reconstruction trial, the characteristic
function is evaluated simultaneously at all $25$ frequencies. A trial is
considered successful only when
\begin{equation}
    \max_{j=1,\ldots,25}
    \left|
    \widetilde{\lambda}(\beta_j)
    -
    \lambda(\beta_j)
    \right|
    \leq\epsilon .
\end{equation}
Thus, the reconstruction error criterion must be satisfied
simultaneously at every sampled frequency, including the endpoint
$\beta_0$. The minimum sample number is determined by a binary search until the success probability reaches
$2/3$. The full-bandwidth threshold determination is repeated $50$ times with
new random selections of the reconstruction datasets and sampled
frequencies. The reported sample complexity is the mean of these
threshold estimates, with their standard deviation used to quantify the
variation arising from the random sampling procedure. The resulting full-bandwidth reconstruction sample complexity is
shown in Fig.~\ref{fig:SM_reconstruction}(d).

Finally, to visualize how reconstruction difficulty varies continuously
across the accessible bandwidth, we evaluate the reconstruction success
probability pointwise as a function of $\beta$. Here $N$ is fixed to the
sample number predicted by the Hoeffding bound for the full-bandwidth
reconstruction protocol at the corresponding squeezing level (7.69 dB). The
success probability is evaluated at $100$ values of $\beta$ spanning
$[0,\beta_0]$. Unlike the simultaneous criterion above, this calculation
evaluates the reconstruction criterion independently at each value of
$\beta$. The resulting dependence is shown in
Fig.~\ref{fig:SM_reconstruction}(c), illustrating the increasing
difficulty of reconstruction towards the edge of the accessible
bandwidth and the improvement obtained with stronger squeezing.

\newpage

\renewcommand{\thesection}{\Alph{section}}
\renewcommand{\thesubsection}{\thesection\arabic{subsection}}
\counterwithout{equation}{section}
\renewcommand{\theequation}{A\arabic{equation}}
\appendix

\section{Rényi Divergence and Binary Hypothesis Testing}
\label{app:renyi-discrimination}

We collect here the information-theoretic relations used in the proof
of Theorem~1. For two classical probability distributions $P$ and $Q$,
the order-two Rényi divergence is related to the Pearson
$\chi^2$ divergence by
\begin{equation}
    D_2(P\Vert Q)
    =
    \log\!\left[1+\chi^2(P\Vert Q)\right],
    \qquad
    \chi^2(P\Vert Q)
    :=
    \sum_x\frac{(P(x)-Q(x))^2}{Q(x)}.
    \label{eq:app-classical-D2}
\end{equation}
The Cauchy--Schwarz inequality gives the total variation distance
\begin{equation}
    \operatorname{TVD}(P,Q)
    \leq
    \frac12\sqrt{\chi^2(P\Vert Q)},
    \label{eq:app-TV-chi}
\end{equation}
while, for equal priors,
\begin{equation}
    P_{\mathrm{succ}}^{\mathrm{opt}}
    =
    \frac12\left[1+\operatorname{TVD}(P,Q)\right].
\end{equation}
Thus, achieving $P_{\mathrm{succ}}^{\mathrm{opt}}\geq1-\delta$ requires
\begin{equation}
    \operatorname{TVD}(P,Q)\geq1-2\delta,
\end{equation}
and hence
\begin{equation}
    D_2(P\Vert Q)
    \geq
    \log\!\left[1+4(1-2\delta)^2\right].
    \label{eq:classicalD2}
\end{equation}

For quantum states, we use the sandwiched Rényi divergence \cite{muller2013quantum}
\begin{equation}
    \widetilde D_\alpha(\rho\Vert\sigma)
    :=
    \frac{1}{\alpha-1}
    \log
    \operatorname{Tr}
    \left[
        \left(
            \sigma^{\frac{1-\alpha}{2\alpha}}
            \rho
            \sigma^{\frac{1-\alpha}{2\alpha}}
        \right)^\alpha
    \right],
\end{equation}
which for $\alpha=2$ becomes
\begin{equation}
    \widetilde D_2(\rho\Vert\sigma)
    =
    \log
    \operatorname{Tr}
    \left[
        \left(
            \sigma^{-1/4}
            \rho
            \sigma^{-1/4}
        \right)^2
    \right].
    \label{eq:app-quantum-D2}
\end{equation}
The sandwiched Rényi divergence is additive on tensor-product states
and obeys the data-processing inequality under a POVM:
\begin{equation}
    \widetilde D_2
    \left(
        \rho^{\otimes N}\middle\Vert\sigma^{\otimes N}
    \right)
    =
    N\widetilde D_2(\rho\Vert\sigma),
    \qquad
    D_2(P\Vert Q)
    \leq
    \widetilde D_2(\rho\Vert\sigma),
    \label{eq:app-additivity}
\end{equation}
where $P$ and $Q$ are the corresponding measurement outcome
distributions. Therefore,
\begin{equation}
    N\widetilde D_2(\rho_1\Vert\rho_0)
    \geq
    \log\!\left[1+4(1-2\delta)^2\right].
    \label{eq:quantumD2}
\end{equation}
For $\delta=1/3$,
\begin{equation}
    N\widetilde D_2(\rho_1\Vert\rho_0)
    \geq
    \log\!\left(\frac{13}{9}\right).
\end{equation}

\section{Thermal-State Mapping}
\label{app:thermal-mapping}

We derive the thermal representation of the Gaussian state $\rho_0$
used in the proof. In the $(\hat x,\hat K)$ basis, with
\begin{equation}
    [\hat x,\hat K]=i,
\end{equation}
the covariance matrix is
\begin{equation}
    \Gamma_0
    =
    \operatorname{diag}
    \left(
        V,\frac{1}{4a}
    \right),
    \qquad
    a:=V_r.
\end{equation}
Its symplectic eigenvalue is
\begin{equation}
    \nu
    =
    \sqrt{\det\Gamma_0}
    =
    \frac12\sqrt{\frac Va}.
    \label{eq:app-nu}
\end{equation}

A centered one-mode Gaussian state can be transformed by a Gaussian
unitary into a thermal state with covariance matrix $\nu I$ \cite{weedbrook2012gaussian}. Thus,
for a suitable bosonic mode $\hat b$,
\begin{equation}
    \rho_0
    =
    (1-q)q^{\hat n},
    \qquad
    \hat n=\hat b^\dagger\hat b,
\end{equation}
where
\begin{equation}
    q
    =
    \frac{2\nu-1}{2\nu+1}
    =
    \frac{\sqrt{V/a}-1}
         {\sqrt{V/a}+1}.
    \label{eq:app-q}
\end{equation}

The required Gaussian transformation is a squeezing transformation.
Writing
\begin{equation}
    \hat b
    =
    \hat a\cosh\tilde r
    -
    \hat a^\dagger\sinh\tilde r,
\end{equation}
the squeezing parameter is
\begin{equation}
    \tilde r
    =
    \frac14\log(aV).
    \label{eq:app-tilde-r}
\end{equation}
Under the same transformation,
\begin{equation}
    \kappa\hat K
    \longmapsto
    z(\hat b+\hat b^\dagger),
    \qquad
    z^2
    =
    \frac{\kappa^2\nu}{2V},
    \label{eq:app-z}
\end{equation}
so that
\begin{equation}
    e^{\kappa\hat K}
    \longmapsto
    e^{z(\hat b+\hat b^\dagger)}.
    \label{eq:app-exponential-map}
\end{equation}

Using Eq.~\eqref{eq:app-q} together with $V=V_r+S$ gives
\begin{equation}
    q^{-1/2}-q^{1/2}
    =
    2\sqrt{\frac{V_r}{S}}.
    \label{eq:app-q-difference}
\end{equation}

\section{Evaluation of the Gaussian Trace}
\label{app:BCH-trace}

From Eq.~\eqref{eq:arb-similarity},
\begin{equation}
    X_s
    =
    c\,e^{s\kappa\hat K}
    \rho_0
    e^{-s\kappa\hat K}.
\end{equation}
In the thermal basis Eq.~\eqref{eq:arb-Tst} becomes,
\begin{align}
T_{st}
&=
c^2(1-q)
\operatorname{Tr}
\Big[
e^{sz(\hat b+\hat b^\dagger)}
q^{\hat n}
e^{-sz(\hat b+\hat b^\dagger)}
(q^{\hat n})^{-1/2}
\notag\\
&\hspace{2.5cm}\times
e^{tz(\hat b+\hat b^\dagger)}
q^{\hat n}
e^{-tz(\hat b+\hat b^\dagger)}
(q^{\hat n})^{-1/2}
\Big].
\label{eq:app-trace-start}
\end{align}

We use
\begin{equation}
    q^{\hat n}\hat bq^{-\hat n}
    =
    q^{-1}\hat b,
    \qquad
    q^{\hat n}\hat b^\dagger q^{-\hat n}
    =
    q\hat b^\dagger,
    \label{eq:app-q-conjugation}
\end{equation}
which implies
\begin{equation}
    q^{\hat n}
    e^{u(\hat b+\hat b^\dagger)}
    q^{-\hat n}
    =
    e^{u(q^{-1}\hat b+q\hat b^\dagger)}.
    \label{eq:app-q-exp}
\end{equation}
Equation~\eqref{eq:app-trace-start} can therefore be rearranged as
\begin{align}
T_{st}
&=
c^2(1-q)
\operatorname{Tr}
\Big[
e^{sz(\hat b+\hat b^\dagger)}
e^{-sz(q^{-1}\hat b+q\hat b^\dagger)}
\notag\\
&\qquad\times
e^{tz(q^{-1/2}\hat b+q^{1/2}\hat b^\dagger)}
e^{-tz(q^{-3/2}\hat b+q^{3/2}\hat b^\dagger)}
q^{\hat n}
\Big].
\label{eq:app-four-exponentials}
\end{align}

Define
\begin{equation}
    L_j=u_j\hat b+v_j\hat b^\dagger.
\end{equation}
Since
\begin{equation}
    [L_i,L_j]
    =
    u_iv_j-v_iu_j
\end{equation}
is a scalar, the BCH formula terminates after the first commutator:
\begin{equation}
    \prod_j e^{L_j}
    =
    \exp\!\left[
        \frac12
        \sum_{i<j}
        (u_iv_j-v_iu_j)
    \right]
    e^{U\hat b+V\hat b^\dagger},
    \qquad
    U=\sum_j u_j,\quad V=\sum_jv_j.
    \label{eq:app-BCH}
\end{equation}
For the four exponentials in Eq.~\eqref{eq:app-four-exponentials},
\begin{equation}
(u_j,v_j)
=
z
\left\{
(s,s),
(-sq^{-1},-sq),
(tq^{-1/2},tq^{1/2}),
(-tq^{-3/2},-tq^{3/2})
\right\}.
\label{eq:app-coefficients}
\end{equation}

The thermal expectation of a linear exponential is
\begin{equation}
    \operatorname{Tr}
    \left[
        \rho_0 e^{U\hat b+V\hat b^\dagger}
    \right]
    =
    \exp\!\left[
        \left(\bar n+\frac12\right)UV
    \right],
    \qquad
    \bar n=\frac{q}{1-q}.
    \label{eq:app-thermal-characteristic}
\end{equation}
Combining this with Eq.~\eqref{eq:app-BCH} gives
\begin{align}
\operatorname{Tr}
\left[
\rho_0\prod_{j=1}^4e^{L_j}
\right]
=
\exp\Bigg[
&
\frac12\sum_{i<j}(u_iv_j-v_iu_j)
\notag\\
&+
\left(\bar n+\frac12\right)
\left(\sum_j u_j\right)
\left(\sum_jv_j\right)
\Bigg].
\label{eq:app-general-BCH-trace}
\end{align}

Substituting Eq.~\eqref{eq:app-coefficients} into the exponent and
simplifying yields
\begin{equation}
    \frac12\sum_{i<j}(u_iv_j-v_iu_j)
    +
    \left(\bar n+\frac12\right)
    \left(\sum_j u_j\right)
    \left(\sum_jv_j\right)
    =
    -2stz^2
    \left(q^{-1/2}-q^{1/2}\right).
\end{equation}
Hence
\begin{equation}
    T_{st}
    =
    c^2
    \exp\!\left[
        -2stz^2
        \left(q^{-1/2}-q^{1/2}\right)
    \right].
\end{equation}
Defining
\begin{equation}
    M
    :=
    2z^2
    \left(q^{-1/2}-q^{1/2}\right),
    \label{eq:app-M}
\end{equation}
we obtain the compact result
\begin{equation}
    T_{st}=c^2e^{-stM}.
    \label{eq:app-Tst}
\end{equation}

Using
\begin{equation}
    z^2=\frac{\kappa^2\nu}{2V},
    \qquad
    \kappa=S\omega,
    \qquad
    \nu=\frac12\sqrt{\frac Va},
\end{equation}
and Eq.~\eqref{eq:app-q-difference}, we find
\begin{equation}
    M
    =
    S\omega^2\sqrt{\frac SV}
    =
    \frac{S^{3/2}\omega^2}{\sqrt V}.
    \label{eq:app-M-final}
\end{equation}

\subsection{Quadratic Form in $\Delta$}
\label{app:quadratic}

From
\begin{equation}
    \Delta=-2i\epsilon_0(X_+-X_-),
\end{equation}
we have
\begin{align}
\operatorname{Tr}
\left[
\Delta\rho_0^{-1/2}
\Delta\rho_0^{-1/2}
\right]
&=
-4\epsilon_0^2
\left(
T_{++}-T_{+-}-T_{-+}+T_{--}
\right).
\label{eq:app-delta-expanded}
\end{align}
Using Eq.~\eqref{eq:app-Tst}, the four terms are
\begin{equation}
    T_{++}=T_{--}=c^2e^{-M},
    \qquad
    T_{+-}=T_{-+}=c^2e^{M}.
\end{equation}
Therefore,
\begin{align}
T_{++}-T_{+-}-T_{-+}+T_{--}
&=
2c^2e^{-M}-2c^2e^M
\notag\\
&=
-4c^2\sinh M,
\end{align}
and hence
\begin{equation}
\operatorname{Tr}
\left[
\Delta\rho_0^{-1/2}
\Delta\rho_0^{-1/2}
\right]
=
16\epsilon_0^2c^2\sinh M.
\label{eq:app-delta-final}
\end{equation}

Finally, using
\begin{equation}
    c^2=e^{-S\omega^2},
    \qquad
    M=S\omega^2\sqrt{\frac SV},
\end{equation}
we obtain
\begin{equation}
    \boxed{
    \mathcal R
    =
    e^{-S\omega^2}
    \sinh\!\left(
        S\omega^2\sqrt{\frac SV}
    \right)
    }.
    \label{eq:app-R-final}
\end{equation}
%\stopcontents[supp]
\newpage

%\bibliographystyle{unsrt}
%\bibliography{apssamp}

\end{document}